\documentclass[10pt]{iopart}
\usepackage{graphicx,epsfig}
\usepackage{bm}
\usepackage{iopams}
\newcommand{\ba}{\begin{eqnarray}}
\newcommand{\ea}{\end{eqnarray}}
\newcommand{\bse}{\numparts}
\newcommand{\ese}{\endnumparts}

\newcommand{\bbq}{\begin{quote}}
\newcommand{\eeq}{\end{quote}}
\newcommand{\tbb}{t_{\textrm{\tiny{bb}}}}
\newcommand{\tcoll}{t_{\textrm{\tiny{coll}}}}
\newcommand{\tmax}{t_{\textrm{\tiny{max}}}}
\newcommand{\Lmax}{L_{\textrm{\tiny{max}}}}
\newcommand{\RR}{{}^3{\cal{R}}}

\newcommand{\T}{{}^3{\cal{T}}}

\newcommand{\EE}{{\cal{E}}}
\newcommand{\FF}{{\cal{F}}}

\newcommand{\VV}{{\cal{V}}}
\newcommand{\HH}{{\cal{H}}}

\newcommand{\hOm}{\hat\Omega}
\newcommand{\hOmi}{\hat\Omega_i}

\newcommand{\Dih}{\delta_i^{(\HH)}}
\newcommand{\Dim}{\delta_i^{(m)}}
\newcommand{\Dik}{\delta_i^{(k)}}

\newcommand{\Da}{\delta^{(A)}}

\newcommand{\Dh}{\delta^{(\HH)}}

\newcommand{\Dm}{\delta^{(m)}}

\newcommand{\Dk}{\delta^{(k)}}

\newcommand{\dd}{{\rm{d}}}
\newcommand{\rtv}{r_{\rm{tv}}}
\newcommand{\ltv}{\ell_{\rm{tv}}}

\begin{document}


\title[Evolution of radial profiles in regular LTB dust models.]{Evolution of radial profiles in regular Lema\^{\i}tre--Tolman--Bondi dust models.} 
\author{ Roberto A. Sussman$^\ddagger$}
\address{
$^\ddagger$Instituto de Ciencias Nucleares, Universidad Nacional Aut\'onoma de M\'exico (ICN-UNAM),
A. P. 70--543, 04510 M\'exico D. F., M\'exico. }
\ead{sussman@nucleares.unam.mx}
\date{\today}
\begin{abstract} We undertake a comprehensive and rigorous analytic study of the evolution of radial profiles of covariant scalars in regular Lema\^\i tre--Tolman--Bondi dust models. We consider specifically the phenomenon of ``profile inversions'' in which an initial clump profile of density, spatial curvature or the expansion scalar, might evolve into a void profile (and vice versa).  Previous work in the literature on models with density void profiles and/or allowing for density profile inversions is given full generalization, with some erroneous results corrected. We prove rigorously that if an evolution without shell crossings is assumed, then only the `clump to void' inversion can occur in density profiles, and only in hyperbolic models or regions with negative spatial curvature. The profiles of spatial curvature follow similar patterns as those of the density, with `clump to void' inversions only possible for hyperbolic models or regions. However, profiles of the expansion scalar are less restrictive, with profile inversions necessarily taking place in elliptic models.  We also examine radial profiles in special LTB configurations:  closed elliptic models, models with a simultaneous big bang singularity, as well as a locally collapsing elliptic region surrounded by an expanding hyperbolic background. The general analytic statements that we obtain allow for setting up the right initial conditions to construct fully regular LTB models with any specific qualitative requirements for the profiles of all scalars and their time evolution. The results presented can be very useful in guiding future numerical work on these models and in revising previous analytic work on all their applications.                            
\end{abstract}

\maketitle
\section{Introduction.}

\footnote{In order to motivate the reading of this long article, we have concentrated most of the background material in the Appendices and have written a proper summary of our results and their implications in the final section (section 11). Readers eager to know these results without going through the technical detail are advised to go directly to this section.} 
The well known spherically symmetric LTB dust models \cite{LTB} are among the most useful known exact solutions of Einstein's equations. There is an extensive literature (see \cite{kras1,kras2} for comprehensive reviews) using these models, as they allow for probing non--linear effects in inhomogeneous sources by means of analytic and/or tractable numeric solutions. In most applications LTB dust solutions lead to simple but effective toy models of cosmological inhomogeneities, as for example in the series of papers \cite{KH1,KH2,KH3,KH4,BKH}. The models are also useful in other theoretical contexts, such as censorship of singularities \cite{sscoll,joshi} and even quantum gravity \cite{quantum}. More recently, LTB models have been extensively used in the widespread effort to explain cosmic dynamics without resorting to an elusive source like dark energy, either by fitting observations \cite{LTB1,LTB2,LTBkolb,num1,num2}, or in the context of scalar averaging of inhomogeneities \cite{LTBfin,LTBchin,LTBave1,LTBave2,LTBave3} (see \cite{celerier,ave_review} for comprehensive reviews on these applications). For a novel theoretical approach to the dynamics of these models see \cite{wainwright}.

Since large scale cosmic structure is dominated by large voids, the study and understanding of the spatial profiles of density and velocity in inhomogeneities is a relevant theoretical and practical issue that could provide important clues on the formation of these voids as part of the dynamics of cosmic sources. While LTB models provide an idealized description of cosmic inhomogeneities, a proper understanding (even qualitative) of radial profiles of scalars in these models still has a significant potential in astrophysical and cosmological implications in this context \cite{kras1,kras2} (see \cite{CBK} for an update). In particular, knowing how these radial profiles evolve is extremely useful when applying to these models a formalism of scalar averaging of inhomogeneities, such as Buchert's formalism \cite{LTBave2,LTBave3}. 

An important theoretical issue in looking at density radial profiles is the possibility that a concavity change could occur under regular conditions (particularly around a symmetry center). This implies that a given initial density profile with a ``clump'' (over--density) form could evolve into a ``void'' (under--density) profile, or vice versa, a phenomenon that is usually denoted by a ``clump to void'' or ``void to clump'' density profile inversion. In a well known article Mustapha and Hellaby \cite{mushel} claimed to have furnished proof for the existence of this profile inversion, and to have found the conditions in which it occurs. However, since the conditions provided by these authors are too complicated (involving second and third order radial derivatives), their proof (their section 5) is based on restrictive assumptions, and thus is not applicable to generic LTB models. In the end, they discuss the observational implications of these profile inversion and furnish basically an empiric proof of their existence by looking at specific models and numeric examples where it occurs.     

Following a completely different methodology from that of Mustapha and Hellaby, we correct, extend and fully generalize their work in this article. By considering fully generic and regular LTB models, we examine the evolution of the radial profiles of the basic covariant scalars: rest mass density, $\rho$, spatial curvature, $\RR$, and Hubble expansion scalar, $\Theta$, aiming  specifically at dealing in full generality with the following issues:\, (i) to provide a precise characterization of radial profiles as ``clumps'' or ``voids'', and (ii) to study in full rigor the conditions that allow for the existence of profile inversions (assuming full regularity). As the reader can find out by going through the article (or by looking at the summary of results in section 11), we do find fully analytic and rigorous conditions for the existence of clump/void radial profiles and their possible inversions for regular LTB models in general. 

An interesting and similar approach to the study of the evolution of radial profiles is found in the articles by Krasinski and Hellaby \cite{KH1,KH4} (see the review in \cite{kras2}). These authors find the conditions for the existence of a unique LTB model, such that a given density or ``velocity'' profile at a given time constant hypersurface can be ``mapped'' to any other profile at a second hypersurface. While there is an obvious intersection and complementarity between the results of this article and those of \cite{KH1,KH4}, our methodology is different and (in our opinion) less restrictive, as we only require generic initial data to be specified at a single time constant hypersurface that can be considered a Cauchy surface in the context of an initial value problem.

A summary of articles that have examined LTB models with void profiles in a perturbative and non--perturbative context can be found in the extensive review of this literature (previous to 1994) given in section 3.1 of \cite{kras1}, and in the update in section 18.5 of \cite{kras2}. Some authors \cite{occhio81,occhio83} have examined non--perturbative void profiles numerically in the context of structure formation scenarios (though, from our results in section 6, the elliptic models in \cite{occhio81} cannot be free from shell crossings). In some suggested configurations the void is modeled as an under--dense FLRW region, which is connected to a cosmic ``background'' through a section of a closed elliptic model \cite{cham,sato}. In other articles \cite{boncham1,boncham2,meszaros}, the void is an LTB region containing a center, matched at a fixed comoving radius to another LTB region (the ``envelope''), which in turn is matched to a FLRW ``exterior''. Although these models are rather artificial, their results agree with those of this article. 

The contents of the article is summarized in the remaining of this section. Section 2 contains the bare basic background material needed to understand the article. The known analytic solutions of the field equations (given in Appendix A), which are determined by a preferred set of free functions ($M,\,E,\,\tbb$), have become the standard formulation in practically all theoretical, empiric and in numeric work on these models. While this standard parametrization is adequate and works in practice, we utilize an alternative set of covariant quasi--local scalars that are more suitable for our purpose (and that can be very convenient for numerical work, as well as a better understanding of several theoretical and practical issues)  \cite{sussQL1,sussQL,suss08,suss09,suss10a,suss10b}. We also express these scalars and their fluctuations as quantities that are scaled with respect to their initial values in the framework of an initial value problem \cite{suss10a}. 

Extra background material, which is important as well but can be distracting if it appears in the main body of the article, has been placed in Appendices B, C and D. For example, the parametrization of the known analytic solutions in terms of the new variables and its initial value formulation (Appendix B), the Hellaby--Lake conditions \cite{suss10a,HLconds,ltbstuff} that guarantee an evolution free from shell crossings (Appendix C), and a discussion (Appendix D) of the relation between the radial coordinate and the proper radial length, which is the affine parameter along radial rays \cite{suss10b}. This is important since we are looking at the behavior of covariant scalars in the radial direction. 

We provide in section 3 a precise and rigorous characterization of the ``clump'' or ``void'' nature of a radial profile of a scalar $A$ in terms of the sign of the  gradient $A'$ (monotonicity of $A$), which leads to the definition of a ``turning value'' (to be denoted ``TV of $A$'') as the value $r=\rtv$ where this gradient vanishes. In section 4 we examine the compatibility between initial clump/void profiles and the Hellaby--Lake conditions, while the concept of ``profile inversion'' is defined in section 5 in terms of the vanishing of the fluctuations defined in section 2, something which can occur with or without the existence of TV's.  In section 6 we examine in detail density radial profiles and provide necessary and sufficient conditions for density profile inversions for parabolic, hyperbolic and open elliptic models or regions. We remark that the locus where the density fluctuation vanishes when there are density TV's corresponds to the ``density wave'' (see section 3.1 of \cite{kras1}). The profiles and profile inversions for spatial curvature and the expansion scalar are respectively examined in sections 7 and 8. The case of closed elliptic models is considered in section 9, as for these models the spherical topology of the hypersurfaces of constant time introduces a TV of the area distance $R$. In section 10 we look at radial profiles and profile inversions in special LTB configurations: simultaneous big--bang singularity, LTB models or regions not containing symmetry centers and configurations constructed by glueing LTB regions with ``mixed'' kinematics (elliptic, parabolic and hyperbolic regions, see \cite{ltbstuff}), and specially the mixed configuration made by an elliptic ``interior'' region surrounded by a hyperbolic ``exterior''.  Section 11 provides a summary and final discussion.      

Appendix E describes an alternative parametrization of the models in terms of quantities that can be related to the Omega and a Hubble parameters of a FLRW cosmology in the homogeneous limit. It is straightforward to translate the results of this article in terms of these parameters, which have been used in various articles \cite{num1,LTBfin,suss08} describing cosmological inhomogeneities in various contexts (see also \cite{suss10a}). Appendix F provides the explicit expressions for the basic covariant scalars that characterize LTB models.

\section{LTB models, kinematic classes and a fluid flow time slicing.}

\footnote{This section provides the minimal background material to make this article as self--contained as possible. The reader is advised to consult references \cite{sussQL1,suss09,suss10a} for details on the quasi--local scalar representation of LTB models. Necessary background material is also found in the Appendices.}  
LTB dust models in their conventional variables are characterized by the following metric and field equations $G^{ab}=\kappa\rho u^au^b$
\begin{equation} ds^2=-c^2dt^2+\frac{R'{}^2}{1+E}\,dr^2+R^2(d\theta^2+ 
\sin^2\theta d\varphi^2),\label{LTB1} \end{equation}
\ba \dot R^2 &=& \frac{2M}{R} +E,\label{fieldeq1}\\
 2M' &=& \kappa \rho \,R^2 \,R',\label{fieldeq2}\ea
where $u^a=\delta^a_0$,\, $\rho$ is the rest--mass density, $\kappa=8\pi G/c^2$,\, $E=E(r),\,M=M(r),\,R=R(ct,r)$, while $R'=\partial R/\partial r$ and  $\dot R=u^a\nabla_a R=\partial R/\partial (ct)$. 

The analytic solutions of (\ref{fieldeq1}) are given in Appendix A as parametric forms involving $R,\,M,\,E$ and a third function $\tbb(r)$ (the ``big bang time''). It is common usage in the literature  (see \cite{kras1,kras2,suss10a,ltbstuff}) to classify these solutions in ``kinematic equivalence classes'' given by the sign of $E$, which determines the existence of a zero of $\dot R^2$. Since $E=E(r)$, the sign of this function can be, either the same in the full range of $r$, in which case we have LTB models of a given kinematic class, or it can change sign in specific ranges of $r$, defining LTB models with regions of different kinematic class (see \cite{ltbstuff}). These kinematic classes are 
\bse\ba E = 0,\qquad \hbox{Parabolic models or regions}\label{par}\\
 E \geq 0,\qquad \hbox{Hyperbolic models or regions}\label{hyp}\\
  E \leq  0,\qquad \hbox{Elliptic models or regions}\label{ell} \ea\ese
where the equal sign in (\ref{hyp}) and (\ref{ell}) holds only in a symmetry center.     

\subsection{Covariant quasi--local scalars.}

The normal geodesic 4--velocity in (\ref{LTB1}) defines a natural time slicing in which the space slices are the 3--dimensional Riemannian hypersurfaces $\T[t]$, orthogonal to $u^a$, with metric $h_{ab}=u_au_b+g_{ab}$, and marked by arbitrary constant values of $t$. 
All radial rays in a given $\T[t]$ are diffeomorphic to each other and to the real line $\mathbb{R}$ (or to continuous subsets of $\mathbb{R}$ if there are symmetry centers). Hence, we will consider every LTB scalar function as equivalent, under the time slicing given by $u^a$, to a one parameter family of real valued functions $A[t]: \mathbb{R}\to \mathbb{R}$ so that $A[t](r)=A(t,r)$.

An LTB model is uniquely determined by a representation of local fluid flow covariant scalars $A$ (see Appendix F), though alternative scalar representations can be defined. For every scalar function $A$ in LTB models admitting (at least) a symmetry center ($r=0$), we define its quasi--local dual $A_q$ as the family of real valued functions $A_q[t]: \mathbb{R}^+\to \mathbb{R}$ given by
\footnote{While we have assumed that the integral in (\ref{aveq_def}) is evaluated for a fully regular $\T[t]$, it is straightforward to generalize this definition to hypersurfaces that intersect singularities and to models without symmetry centers (see Appendix A--5 of \cite{suss10a}). In order to simplify the notation we will omit henceforth the symbol $[t]$ attached to scalar functions unless it is needed for clarity.}
\begin{equation}  A_q=\frac{\int_0^{r}{A\,\FF\,\dd\VV_p}}{\int_0^{r}{\FF\,\dd\VV_p}}=\frac{\int_0^{r}{A R^2 R'\,\dd x}}{\int_0^{r}{R^2 R'\dd x}},\label{aveq_def}\end{equation}
where the integration is along arbirtary slices $\T[t]$,\,with $\FF\equiv (1+E)^{1/2}$,\, $\dd\VV_p=\sqrt{h_{ab}}\dd r\dd\theta\dd\varphi$, and we are using the notation $\int_0^r{... \dd x}=\int_{x=0}^{x=r}{... \dd x}$. 

The definition of quasi--local scalars (\ref{aveq_def}) leads in a natural manner to an initial value parametrization of LTB models, so that all quantities can be scaled in terms of their value at a fiducial (or ``initial'') slice $\T_i\equiv \T[t_i]$, where $t=t_i$ is arbitrary. Hence, the subindex ${}_i$ will denote henceforth ``initial value functions'', which will be understood to be scalar functions evaluated at $t=t_i$. This scaling of quantities to $\T_i$ suggests introducing 
\begin{equation} L \equiv \frac{R}{R_i},\label{Ldef}\end{equation}
so that dependence on $R$ becomes dependence on $L$ as a sort of dimensionless scale factor. By introducing the following definitions (which simplify notation)
\begin{equation}\fl 2m\equiv \frac{\kappa}{3}\,\rho,\quad 2m_q\equiv \frac{\kappa}{3}\,\rho_q,\qquad k\equiv \frac{\RR}{6},\quad k_q\equiv \frac{\RR_q}{6},\qquad \HH\equiv\frac{\Theta}{3},\quad \HH_q\equiv\frac{\Theta_q}{3},\label{mkHdefs}\end{equation}
it is straightforward to obtain from (\ref{fieldeq1}), (\ref{fieldeq2}), (\ref{ThetaRR}), (\ref{aveq_def}), (\ref{Ldef}) and (\ref{mkHdefs}) the quasi--local density, spatial curvature and expansion scalar 
\ba m_q = \frac{m_{qi}}{L^3} = \frac{M}{R^3},\label{mq}\\
 k_q = \frac{k_{qi}}{L^2} =-\frac{E}{R^2},\label{kq}\\
 \HH_q^2 = \frac{\dot L^2}{L^2}=\frac{\dot R^2}{R^2}=2m_q-k_q=\frac{2m_{qi}-k_{qi}L}{L^3},\label{Hq}\ea     
so that the conventional free functions $M$ and $E$ are expressible in terms of initial value functions by
\begin{equation} M = m_{qi}\,R_i^3,\qquad E = -k_{qi}\, R_i^2.\label{ME} \end{equation}

\subsection{Fluctuations.}

The quasi--local scalars defined by (\ref{aveq_def}) comply with the following properties
\bse\ba    
 A_q'=(A_q)' = \frac{3R'}{R}\,\left[A-A_q\,\right],\label{propq2}\\
A(r) - A_q(r) = \frac{1}{R^3(r)}\int_0^r{A' \,R^3 \,\dd x}.\label{propq3}\ea\ese
Hence, given the pair of scalars $\{A,\,A_q\}$, we define their relative fluctuations as
\begin{equation} \Da \equiv \frac{A-A_q}{A_q}=\frac{A'_q/A_q}{3R'/R} =\frac{1}{A_q(r) R^3(r)}\int_0^r{A'\,R^3\,\dd x},\label{Dadef}\end{equation}
where we used (\ref{propq2}) and (\ref{propq3}). The local density, spatial curvature (Ricci scalar of the $\T[t]$) and expansion scalar $\Theta =\tilde\nabla_a u^a$, given by (\ref{fieldeq2}) and (\ref{ThetaRR}), are then expressible in terms of the fluctuations (\ref{Dadef}) or their gradients  (from (\ref{propq2})): 
\bse\ba  \fl m=m_q\,\left[1+\Dm\right] = m_q+\frac{m'_q}{3R'/R},\label{qltransfm}\\
\fl  k=k_q\,\left[1+\Dk\right] = k_q+\frac{k'_q}{3R'/R}.\label{qltransfk}\\
\fl  \HH=\HH_q\,\left[1+\Dh\right] = \HH_q+\frac{\HH'_q}{3R'/R},\label{qltransfH} \ea\ese
The remaining local fluid flow scalars in (\ref{SigEE}) and (\ref{SigEE1}), associated with the shear and electric Weyl tensors, follow as  
\begin{equation}  
\fl \Sigma = -\HH_q\,\Dh = -\frac{\HH'_q}{3R'/R},\qquad 
\EE=-m_q\,\Dm =-\frac{m'_q}{3R'/R},\label{qltransfSE}
\end{equation}
By introducing the scale factor
\begin{equation}\Gamma \equiv \frac{R'/R}{R'_i/R_i}=1+\frac{L'/L}{R'_i/R_i},\label{Gdef}\end{equation} 
the following scaling laws for the local density and spatial curvature follow readily from (\ref{fieldeq2}), (\ref{kq}), (\ref{mq}) and (\ref{ThetaRR}) as:
\bse\ba  m=\frac{m_{qi}}{L^3}\,[1+\Dm] =\frac{m_i}{L^3\,\Gamma},\label{slaw1}\\
    k=\frac{k_{qi}}{L^2}\,[1+\Dk] = \frac{k_i}{L^2\,\Gamma}\,\left[1+\frac{\Gamma-1}{3\,(1+\Dik)}\right].\label{slaw2}\ea\ese 
Hence, by comparing (\ref{slaw1}) and (\ref{slaw2}) with (\ref{qltransfm}) and (\ref{qltransfk}), we obtain the following scaling laws for the fluctuations $\Dm$ and $\Dk$ 
\bse\ba 1+\Dm=\frac{1+\Dim}{\Gamma},\label{slawDm}\\
\frac{2}{3}+\Dk = \frac{2/3+\Dik}{\Gamma},\label{slawDk}\\
\Dm-\frac{3}{2}\Dk =\frac{\Dim-(3/2)\Dik}{\Gamma},\label{slawD32}\ea\ese  
while the scaling law for $\Dh$ follows from (\ref{Dadef}), (\ref{qltransfH}) and (\ref{qltransfSE}):   
\ba  2\Dh &=& \frac{2m_q\,\Dm-k_q\,\Dk}{2m_q-k_q}=\frac{2m_{qi}\,\Dm-k_{qi}\,L\,\Dk}{2m_{qi}-k_{qi}\,L}\nonumber\\
 &=& \frac{2m_{qi}\,[\Dim+1-\Gamma]-k_{qi}\,L\,[\Dik+\frac{2}{3}(1-\Gamma)]}{[2m_{qi}-k_{qi}L]\,\Gamma},\label{slawDh}\ea
which allows us to obtain scaling laws for the local expansion scalar, $\HH$, and the scalar functions $\Sigma$ and $\EE$.

\subsection{LTB metric and curvature singularities.}

It is important to remark that the scalars $m_q,\,\HH_q,\,k_q$ and their fluctuations are covariant objects, as $M,\,E,\,R,\,\dot R=u^a\nabla_a R$ are invariants in spherically spacetimes \cite{hayward}. 
Given (\ref{Ldef}), (\ref{ME}) and (\ref{Gdef}), the LTB metric (\ref{LTB1}) takes the form
\begin{equation} \dd s^2=-c^2\dd t^2+L^2\left[\frac{\Gamma^2\,{R'_i}^2\,\dd r^2}{1-k_{qi}\,R_i^2}+R_i^2\left(\dd\theta^2+\sin^2\theta\dd \phi^2\right)\right],\label{LTB2}\end{equation}
The Friedman--like equation (\ref{fieldeq1}) now takes the form (\ref{Hq}), whose solutions are summarized in Appendix B and are equivalent to those of (\ref{fieldeq1}) in Appendix A.  However, the solutions of (\ref{Hq}) are expressed in terms of $L$ and the initial value functions $m_{qi},\,k_{qi},\,R_i$, with the ``big bang time'' $\tbb(r)$ given in terms of the latter functions. The analytic solutions of (\ref{Hq}) in Appendix B allow us to compute the following analytic form for $\Gamma$ by implicit radial derivation of (\ref{par2}), (\ref{hypZ2}) and (\ref{ellZ2}), and then using (\ref{Dadef}) and $L'/L=(1-\Gamma)R_i'/R_i$ to eliminate $m'_{qi},k'_{qi}$ and $L'$ in terms of $\Dim,\,\Dik$ and $\Gamma$. The result is
\begin{itemize}
\item {\underline{Parabolic models or regions.}}
\begin{equation} \Gamma = 1+\Dim-\frac{\Dim}{L^{3/2}}.\label{Gp}\end{equation}
\item {\underline{Hyperbolic and elliptic models or regions.}}
\ba  \fl \Gamma = 1+3(\Dim-\Dik)\left(1-\frac{\HH_q}{\HH_{qi}}\right)
-3\HH_q\,c(t-t_i)\,\left(\Dim-\frac{3}{2}\Dik\right),\label{Ghe}\ea
\end{itemize}

\noindent
where $\HH_q$ and $\HH_{qi}$ follow from (\ref{Hq}), while $c(t-t_i)$ is given by (\ref{hypZ2}) and (\ref{ellZ2}). 

By inserting (\ref{Gp}) or (\ref{Ghe}) (depending on the kinematic class) into (\ref{slaw1})--(\ref{slaw2}), (\ref{slawDm})--(\ref{slawDk}) and (\ref{slawDh}) leads to closed analytic expressions for all these scaling laws in terms of $L$ and initial value functions in the context of the initial value approach described in previous subsection. The functional forms of $\Gamma$ and $\HH_q$ in (\ref{Hq}) and (\ref{Gp})--(\ref{Ghe}) allow us to obtain analytic expressions for all the scaling laws of covariant scalars (as functions of $L$ and initial value functions).

The scaling laws (\ref{mq})--(\ref{slawDh}) clearly indicate the existence of two possible curvature singularities whose coordinate locus is
\bse\ba L(t,r) &=& 0,\qquad \hbox{central singularity}\label{Lzero}\\
 \Gamma(t,r) &=& 0 \qquad \hbox{shell crossing singularity}.\label{Gzero}\ea\ese
so that for reasonable initial value functions (bounded and continuous), all scalars $A_q=m_q,\,k_q,\,\HH_q$ diverge as $L\to 0$, whereas local scalars $A=m,\,k,\,\HH$ can also diverge if $\Gamma\to 0$ (even if $L>0$). 

Notice that if $\Gamma>0$, then all scalars $A$ and $A_q$ only diverge at the central singularity $L=0$, which is an intrinsice feature of LTB models. However, there is a shell crossing singularity if $\Gamma\to 0$ for $L>0$, so that all the relative fluctuations $\Da$  and local scalars $A$ diverge while their quasi--local duals $A_q$ remain bounded (with $A_q\ne 0$). This is an obviously unphysical effect of shell crossings that must be avoided. We will denote by ``regular LTB models'' all configurations for which shell crossing singularities are absent, thus complying with
\begin{equation} \Gamma > 0 \quad\forall\;\; (ct,r)\quad\hbox{such that}\quad L>0 \label{noshxG}\end{equation}
In order to test this regularity condition we need to check the sign in the forms of $\Gamma$ in (\ref{Gp}) and (\ref{Ghe}). However, as proven by Hellaby and Lake \cite{HLconds,ltbstuff}, these conditions can be specified in terms of $M,\,E,\,R$ and their gradients, which in terms of our parameters translates into being specified in terms of initial value functions $m_{qi},\,k_{qi}$ and their initial fluctuations $\Dim,\,\Dik$. The Hellaby--Lake conditions are summarized in Appendix C separately for parabolic, hyperbolic and elliptic models or regions. 

\section{Qualitative features of radial profiles of scalars.}

\subsection{The radial coordinate.}

The study of radial profiles of covariant scalars (local or quasi--local or fluctuations) is basically the study of their behavior along radial rays in the hypersurfaces $\T[t]$ orthogonal to $u^a$. Since the radial coordinate has no intrinsic covariant meaning, while the radial rays are spacelike geodesics of the LTB metric \cite{suss10b}, we should use their affine parameter (proper radial length) $\ell$ given by (\ref{elldef}) to probe the radial behavior of covariant scalars. \footnote{See Appendix D for a discussion on the regularity of the proper radial length and the radial coordinate.} Though, unfortunately, the dependence of scalars on $\ell$ is extremely difficult to evaluate, even qualitatively. However, as long as we assume that (\ref{noshxG}) holds (shell crossings are absent) and $r$ is a well defined coordinate so that $R_i$ and $k_{qi}$ comply with (\ref{layer}), the proper length $\ell$ is a monotonously increasing function of $r$. As a consequence, dependence on $\ell$ becomes qualitatively analogous to dependence on $r$ (see \cite{suss10b}), and thus, all results given in terms of $r$ regarding features of radial profiles of scalars (monotonicity, TV's or asymptotics) will be qualitatively analogous and fully equivalent to results given in terms of $\ell$.  We will henceforth assume that (\ref{noshxG}) and (\ref{layer}) hold. 

The fact that the radial coordinate in (\ref{LTB1}) or (\ref{LTB2}) admits an arbitrary rescaling $r=r(\bar r)$ defines a `radial coordinate gauge freedom' that allows us to use any function of $r$ as a radial coordinate (for example setting $M$ as radial coordinate as in \cite{KH1,KH2,KH3,KH4} and many other articles). In particular, the metric form (\ref{LTB2}) suggests using this coordinate gauge freedom to prescribe a functional form for the initial value function $R_i(r)$ \cite{suss09,suss10a,suss10b}. However, the form of $R_i$ is not completely arbitrary: it is restricted by the topology of the space slices $\T[t]$ and by regularity conditions (see Appendix D and also \cite{suss02} and Appendix A3 of \cite{suss10a}). 

\subsection{Monotonicity and characterization of radial profiles as ``clumps'' or ``voids''.}

Besides their role as covariant non--linear perturbations \cite{suss09,suss10a}, the relative fluctuations $\Da$ convey important non--local information on the radial profile of scalars $A$. Consider the radial domain ($t$ constant) 
\begin{equation}\vartheta_c(r)=\{x\,|\,0\leq x \leq r\},\label{vartheta}\end{equation} 
where the lower bound $r=0$ marks a symmetry center. We shall assume henceforth that $R'>0$ holds for every domain of this form. We have then:    

\begin{quote}

\noindent \underline{Lemma 1}. Let $A$ be a smooth integrable scalar function defined in $\vartheta_c(r)$. If $A$ is monotonous $\Rightarrow$ $A_q$ is monotonous, and the following results hold for its quasi--local dual $A_q$ and relative fluctuation $\Da$ defined by (\ref{aveq_def}) and (\ref{Dadef})

\begin{itemize} 

\item $\rm{sign}(A-A_q)=\rm{sign}(A_q\Da)=\rm{sign}(A')$. 

\item If $A\ne 0$, then $\rm{sign}(\Da)=\rm{sign}(A'/A)$.

\end{itemize}

\noindent \underline{Corollary}. The following sign relations hold:
\begin{itemize}
\item If $A\geq 0$, then $A_q\geq A$ when $A'\leq 0$ and $A_q\leq A$ when $A'\geq 0$.  
\item If $A\leq 0$, then $A_q\leq A$ when $A'\leq 0$ and $A_q\geq A$ when $A'\geq 0$.
\end{itemize}
\noindent \underline{Proof.} These results follow directly from (\ref{propq2}), (\ref{propq3}) and (\ref{Dadef}). The converse  statements are not true in general (see the comment to Lemma 3).

\smallskip

\noindent \underline{Comment}. We will consider (unless stated otherwise) domains of the form  $\vartheta_c(r)$ whose lower bound is a symmetry center. However, Lemma 1 and its corollary are valid for any regular radial domain, in particular domains not containing a symmetry center. Such domains arise in either one of the following situations:
\begin{itemize}
\item When looking at a given region in compound configurations made by elliptic, hyperbolic and parabolic regions \cite{ltbstuff} (see section 10).
\item LTB models that do not admit symmetry centers (see section 10).     
\item Radial domains of hypersurfaces $\T[t]$ that intersect a central singularity marked by $ct=c\tbb(r)$ or $ct=c\tcoll(r)$ in the $(ct,r)$ plane (see Appendix A--5 of \cite{suss10a}).
\end{itemize}
In all these cases the domain must be a suitable modification or restriction of $\vartheta_c(r)$. In the case of the intersection of $\T[t]$ with a singularity, the domain can given by $\bar\vartheta(r)=\{x\,|\,r_{\textrm{\tiny{sing}}} < x \leq r\}$, where $r_{\textrm{\tiny{sing}}}$ is the value of the radial coordinate marking the intersection of $c\tbb(r)$ or $c\tcoll(r)$ with a given $\T[t]$. Evidently, $\bar\vartheta(r)$ does not contain a symmetry center.   

\end{quote} 

\noindent
\underline{Comment: clump and void profiles.}\,\, Lemma 1 indicates how, for a monotonous radial profile of $A$ in a domain (\ref{vartheta}), the sign of $\Da$ determines if this profile has the form of a ``clump'' or ``void''. 
\footnote{This characterization becomes more nuanced when dealing with domains not containing a center, and also when $R'=0$ holds without violating regularity conditions (in closed elliptic models in which the $\T[t]$ have $\mathbb{S}^3$ topology). See sections 9 and 10.}
Consider a non--negative quantity such as $2m=\kappa\rho/3$, a density clump ($m'\leq 0$) in a domain $\vartheta_c(r)$ around a symmetry center of an arbitrary hypersurface $\T[t]$ will be characterized by $\Dm\leq 0$, whereas for a void ($m'\geq 0$) we have $\Dm\geq 0$. If we consider the case of negative spatial curvature ($k\leq 0$), then the signs of $k'$ invert:  $k'\geq 0$ for a clump and $k'\leq 0$ for a void, but in either case the sign of $\Da$ remains the same as that for positive quantities: $\Dm\leq 0$ for a clump and $\Dm\geq 0$ for a void. In case there is a zero of $A_q$, making $\Da$ diverge, then the ``clump'' or ``void'' nature of a radial profile should be characterized by the sign of $A-A_q=A_q\Da$.  Since the results of this lemma are valid for an arbitrary $t$, they can be very useful for characterizing the features of initial conditions at a given $t=t_i$ and their evolution for $t\ne t_i$.

\subsection{Turning values (TV's)}

If the radial profile of a scalar in a given $\T[t]$ is not monotonous, then there exists (at least) one turning value of $A$  (a ``TV of $A$''), which is a value $\rtv\in \vartheta_c(r)$ such that $A'(\rtv)=0$. We discuss below the conditions for the existence of a TV.  Let $A$ be a smooth integrable scalar function defined in the domain $\vartheta_c(r)$ in (\ref{vartheta}) along an arbitrary $\T[t]$, then

\begin{quote}

\noindent \underline{Lemma 2}. The necessary and sufficient condition for the existence of a TV of $A_q$ in a hypersurface $\T[t]$ is
\begin{equation}\Da(ct,\rtv)=0,\label{lemma2}\end{equation}
\noindent \underline{Proof.} The result follows directly from (\ref{Dadef}) given the assumption that $R'>0$ holds in every $\vartheta_c(r)$.  Since $A_q=A_q(t,r)$, then the value of $\rtv$ will (in general) change for different $\T[t]$ (see figure \ref{fig1}).   

\smallskip

\noindent \underline{Corollary 1.} If $\rtv\in\vartheta_c(r)$ marks a TV of $A_q$, then $A_q(\rtv)=A(\rtv)$. 

\smallskip

\noindent \underline{Corollary 2.} From section 4 of \cite{suss10b}: if (\ref{layer}) holds, then \;\;$\exists\,\, r=\rtv$ such that $A'(\rtv)=0\;\;\Leftrightarrow\;\;\exists\,\, \ell=\ltv$ such that $\partial A/\partial\ell=0$ at $\ell=\ltv$,  where $\ell$ is proper length (\ref{elldef}) along the $\T[t]$. 

\smallskip

\noindent \underline{Comment 1.} When $A$ is the density, the locus of (\ref{lemma2}) in the $(ct,r)$ plane is the so--called ``{\bf {density wave}}'' mentioned in section 3.1 of \cite{kras1}. 

\smallskip

\noindent \underline{Comment 2.} The case in which $R'=0$ occurs under regular conditions ($\T[t]$ homeomorphic to $\mathbb{S}^3$) is discussed in section 13. Notice that $\Da\to\pm\infty$ with $\Gamma>0$ marks a zero of $A_q$, not (in general) a TV of $A_q$ (see Appendix A4 of \cite{suss10a}).
\\

\noindent \underline{Lemma 3}. The existence of a TV of $A_q$ in $\vartheta_c(r)$ is a sufficient condition for the existence of a TV of $A$. 

\smallskip

\noindent \underline{Proof.} Considering the corollary of the previous lemma and using (\ref{propq3}) we have at $\rtv$
\begin{equation} A(\rtv)-A_q(\rtv)=\int_0^{\rtv}{A'(x)\frac{R^3(x)}{R^3(\rtv)}\dd x}=0.\label{lemma3}\end{equation}
Since $R^3(x)>0$ for $x>0$ and $R^3(\rtv)$ are positive, then for this integral to vanish $A'$ must have a zero in the range $0<x<\rtv$. 

\smallskip

\noindent \underline{Comment.} The converse of this lemma is not true:  a TV of $A$ does not imply a TV of $A_q$ in every domain. This is connected to the fact that a monotonous $A_q$ does not imply (in general) a monotonous $A$ (the converse of Lemma 1). Consider a domain $\vartheta_c(\rtv)$. The TV of $A$ occurs at a value $x=r_1<\rtv$, therefore for values $r_1<x<\rtv$ the gradient $A'$ has already changed signs whereas $A'_q$ has not. However, if we consider domains with $r>\rtv$ and there is a TV of $A$ at $x=r_1$ and $R'>0$ in $\vartheta_c(r)$, then for sufficiently large $r$ there will always be a TV of $A_q$ at $x=\rtv>r_1$.\\

\end{quote}

\noindent
It is clear, as a consequence of the lemmas proven above, that knowledge on the qualitative form of the radial profile of $A_q$ in a given $\T[t]$, {\it{i.e.}} monotonicity, TV's and radial asymptotics (see next section), immediately translates into knowledge on the radial profile of its dual scalar $A$.  As shown in section 2, the quasi--local scalars $m_q,\,k_q,\,\HH_q$ satisfy scaling laws that are less complicated than those of their local counterparts $m,\,k,\,\HH$.  Hence, it is more practical to examine first the radial profiles of the scalars $A_q$ in order to use the lemmas of this section to infer the profiles of the $A$. Finally, we remark that  monotonicity or existence of TV's can change from one $\T[t]$ to the other.

\section{Initial profiles compatible with absence of shell crossings. }

It is worthwhile looking carefully at the types of profiles (clumps or voids) compatible with initial value functions $m_{qi},\,k_{qi}$ (or $\HH_{qi},\,\hOmi$ in Appendix E) that satisfy (\ref{noshxG}), and so assure a time evolution free from shell crossing singularities. For this purpose, we need to examine the allowed signs of $\Dim$ and $\Dik$ involved in the Hellaby--Lake conditions (\ref{noshx_par}), (\ref{noshxGh}) and (\ref{noshxGe}). 

It is evident (from (\ref{Dadef}) and Lemma 1) that clumps or void initial profiles directly correlate with the sigs of the gradients $m'_{qi},\,k'_{qi}$, but it is not so clear how this type of profiles can be prescribed by means of the conventional free functions $M$ and $E$. To appreciate the relation between $\Dm,\,\Dk$ and their initial values $\Dim,\,\Dik$ to the functions $M$ and $E$ and their gradients, we use the scaling laws (\ref{slawDm}) and (\ref{slawDk}) together with (\ref{Dadef}), (\ref{mq}) and (\ref{kq}), leading to the general form of the relation (\ref{MEDi})   
\bse\ba  \frac{M'}{M}=\frac{3R'_i}{R_i}\,\Gamma\,[1+\Dm]\,=\frac{3R'_i}{R_i}\,[1+\Dim],\label{scaling1a}\\
 \frac{E'}{E}=\frac{3R'_i}{R_i}\,\Gamma\,\left[\frac{2}{3}+\Dk\right]=\frac{3R'_i}{R_i}\,\left[\frac{2}{3}+\Dik\right],\label{scaling1b} \ea\ese
From these relations, we can convey the characterization of clump ($\Da_i\leq 0$) or void ($\Da_i\geq 0$) initial profiles to the following constraints on $M$ and $E$ and their gradients:
\bse\ba \fl \hbox{Initial}\,\,m:\qquad \frac{M'/M}{3R'_i/R_i}\leq 1\quad(\hbox{clump})\qquad  \frac{M'/M}{3R'_i/R_i}\geq 1\quad(\hbox{void}),\label{clumpME}\\
\fl \hbox{Initial} \,\, k:\qquad \frac{E'/E}{2R'_i/R_i}\leq 1\quad(\hbox{clump})\qquad  \frac{E'/E}{2R'_i/R_i}\geq 1\quad(\hbox{void}),\label{voidME}\ea\ese
where $E>0$ and $E<0$ hold, respectively, for hyperbolic and elliptic models or regions. 

The parabolic case is trivial. Since fulfillment of (\ref{noshxG}) only requires the Hellaby--Lake condition (\ref{noshx_par}): $-1<\Dim\leq 0$, the initial value functions $m_{qi}$ or $\HH_{qi}$ must have clump profiles, so that $M$ must comply with the restriction in the clump case in (\ref{clumpME}). The hyperbolic and elliptic cases require further examination.

\subsection{Hyperbolic models or regions}

The Hellaby--Lake conditions (\ref{noshxGh}) do not rule out a clump or void profile of $m_{qi}$ and $k_{qi}$. The first two conditions simply place a lower negative bound for $\Dim$ and $\Dik$, hence to get further information we look at $\tbb'\leq 0$ from (\ref{tbbr}) rewritten as
\begin{equation} \alpha\,\Dim+\beta\,\Dik\leq 0,\label{tbbrh}\end{equation}
where\
\bse\ba\fl\alpha=1-\HH_{qi}c(t_i-\tbb)=-\frac{2}{x_i}+\frac{[2+x_i]^{1/2}\hbox{arccosh}(1+x_i)}{x_i^{3/2}},\label{alphah}\\
\fl\beta=\frac{3}{2}\HH_{qi}c(t_i-\tbb)-1=\frac{1}{2}+\frac{3}{x_i}-\frac{3[2+x_i]^{1/2}\hbox{arccosh}(1+x_i)}{2x_i^{3/2}},\label{betah}\\
\fl x_i=\frac{|k_{qi}|}{m_{qi}}.\ea\ese
Since $\alpha\geq 0$ and $\beta\geq 0$ for every choice of initial value functions (either $m_{qi},\,|k_{qi}|$ or $\HH_{qi},\,\hOmi$), then (\ref{tbbrh}) necessarily excludes the possibility that $\Dim\geq 0,\,\,\Dik\geq 0$, and so initial void profiles for density and curvature are incompatible with the fulfillment of (\ref{noshxG}).  On the other hand, sufficient (but not necessary)\footnote{This corrects an error in \cite{suss02}, where these conditions were regarded as necessary and sufficient} conditions for the fulfillment of (\ref{noshxG}) follow by choosing initial density and curvature profiles of clumps:
\begin{equation} -1\leq \Dim\leq 0,\qquad -\frac{2}{3}\leq\Dim\leq 0. \end{equation}
Still, initial density voids $\Dim\geq 0$ are compatible with (\ref{noshxG}) provided initial curvature is a clump: $\Dik\leq 0$. The opposite situation: density clump and curvature void is also compatible with (\ref{noshxG}). In fact, by looking at the functional forms in (\ref{alphah}) and (\ref{betah}), we have $\alpha=\beta$ for $x_i\approx 4.11$ (or $\hOmi\approx 0.327$), with:
\bse\ba \alpha>\beta\quad\hbox{for}\quad 0<x_i<4.11\quad \hbox{or}\quad 0<\hOmi<0.327,\\
 \alpha<\beta\quad\hbox{for}\quad x_i>4.11\quad \hbox{or}\quad 0.327<\hOmi<1,\ea\ese
where $\hOmi=2m_{qi}/\HH_{qi}^2=2/(2+x_i)$ (see Appendix E). Hence, other sufficient (but not necessary) conditions for (\ref{noshxG}) involving initial voids of $m$ and $k$ can be given by selecting $x_i$ to be restricted to specific ranges: 
\bse\ba \fl\Dim\geq 0,\,\Dik\leq0\quad \hbox{with}\quad \Dim\geq|\Dik|\quad\hbox{for}\quad 0<\hOmi<0.327,\\
\fl \Dim\leq 0,\,\Dik\geq0\quad \hbox{with}\quad \Dik\geq|\Dim| \quad \hbox{for}\quad 0.327<\hOmi<1,\ea\ese
In general, given a choice of $m_{qi},\,|k_{qi}|$, a very practical way to test the fulfillment of (\ref{noshxG}) follows by rewriting the Hellaby--Lake conditions (\ref{noshxGh}) as
\begin{equation} \fl \frac{m'_{qi}}{m_{qi}}\geq -\frac{3R'_i}{R_i},\quad \frac{k_{qi}'}{k_{qi}}\geq -\frac{2R'_i}{R_i},\quad m'_{qi}\,x_i\alpha(x_i)+k_{qi}'\,\beta(x_i)\leq 0,\quad k_{qi}\leq 0,\label{noshx_hcon}\end{equation}
where $\alpha$ and $\beta$ are given by (\ref{alphah}) and (\ref{betah}), and $R_i$ can always be specified as a coordinate choice (for example, as in (\ref{rgauge})). The constraint (\ref{noshx_hcon}) can be given in terms of $\hOmi,\,\HH_{qi}$ by means of (\ref{OmD}). In terms of $M$ and $E$, we have $M'\geq 0$ and $E'\geq 0$, as in (\ref{noshx1}). 

It can be argued that choosing the free functions $M,\,E$ and $\tbb$ directly from the Hellaby--Lake conditions, as in (\ref{noshx1}), is easier than testing (\ref{tbbrh}) or (\ref{noshx_hcon}). However, the compensation for going through these more complicated constraints lies in the fact that, for some specific problems, it can be more practical and/or intuitive to select initial $\hOmi$ and $\HH_{qi}$, related to the Omega and Hubble factors (see Appendix E), or an initial density and curvature profile, than functional forms for $M,\,E$ and $\tbb$. 

\subsection{Elliptic models and regions.}

It is evident that the Hellaby--Lake conditions (\ref{noshxGe}) are far more restrictive than in the hyperbolic case (\ref{noshxGh}). In order  to examine the type of initial profiles (clumps or voids) compatible with (\ref{noshxG}), we rewrite the first two conditions in (\ref{noshxGe}) as
\bse\ba \alpha_1\,\Dim-\beta_1\,\Dik\leq 0,\label{tbbre}\\
\alpha_2\,\Dim-\beta_2\,\Dik\geq 0,\label{tcolre}\ea\ese
with $\alpha_1,\,\beta_1$ and  $\alpha_2,\,\beta_2$ given by
\bse\ba \alpha_1=\frac{2}{x_i}-\frac{(2-x_i)^{1/2}\,\hbox{arccos}(1-x_i)}{x_i^{3/2}},\label{alpha1}\\
  \beta_1=\frac{6-x_i}{2x_i}-\frac{3(2-x_i)^{1/2}\,\hbox{arccos}(1-x_i)}{2x_i^{3/2}},\label{beta1}\ea\ese
\bse\ba \alpha_2 &=& \frac{2}{x_i}+\frac{(2-x_i)^{1/2}\,[2\pi-\hbox{arccos}(1-x_i)]}{x_i^{3/2}},\label{alpha2}\\
  \beta_2 &=& \frac{6-x_i}{2x_i}+\frac{3(2-x_i)^{1/2}\,[2\pi-\hbox{arccos}(1-x_i)]}{2x_i^{3/2}},\label{beta2}\ea\ese 
where $0< x_i\leq 2$ (which is equivalent to $\hOmi>1$).  Since $\alpha_1,\,\beta_1$ and  $\alpha_2,\,\beta_2$ are all positive for every choice of initial value functions $m_{qi},\,k_{qi}$ or $\hOmi,\,\HH_{qi}$, we can combine these conditions into a single inequality:
\begin{equation}\frac{\beta_2}{\alpha_2}\Dik\leq \Dim\leq\frac{\beta_1}{\alpha_1}\Dik, \quad\hbox{where}\quad \frac{3}{2}>\frac{\beta_2}{\alpha_2}> 1>\frac{\beta_1}{\alpha_1}>0.\label{noshxce}\end{equation} 
Since this inequality is valid for all initial value functions $m_{qi},\,k_{qi}$ or $\HH_{qi},\,\hOmi$, then it clearly implies the following necessary (but not sufficient) condition for an evolution free from shell crossings complying with (\ref{noshxG})
\begin{equation} \Dim\leq 0,\qquad \Dik\leq 0.\label{noshx_nec2}\end{equation}
Hence, as opposed to the hyperbolic case, the Hellaby--Lake conditions (\ref{noshxGe}) are incompatible with initial voids of density and curvature. The functions $M$ and $E$ must then comply with the clump restrictions in (\ref{clumpME})--(\ref{voidME}). Notice that (\ref{noshxGe}) imposes the restriction $\Dim\geq -1$, but places no lower bounds on a negative $\Dik$.

The constraints (\ref{tbbre})--(\ref{tcolre}) and (\ref{noshxce}) are too complicated for guessing the form of the appropriate initial value functions $m_{qi},\,k_{qi}$ or $\HH_{qi},\,\hOmi$ to fulfill (\ref{noshxGe}). It is more practical (see \cite{suss02}) to test any given ansatz for these functions first on the necessary conditions (\ref{Gmax}), (\ref{noshx_nec1}) and (\ref{noshx_nec2}), and once verifying that these conditions are satisfied, the ansatzes should be tested on (\ref{noshxce}), which can be written in a form similar to (\ref{noshx_hcon}):
\begin{equation} \frac{k'_{qi}}{k_{qi}}\frac{\beta_2}{\alpha_2}\leq \frac{m'_{qi}}{m_{qi}}\leq \frac{k'_{qi}}{k_{qi}}\frac{\beta_1}{\alpha_1}\end{equation}
where $\alpha_1,\,\alpha_2,\,\beta_1,\,\beta_2$ are given by (\ref{alpha1})--(\ref{beta1}) and (\ref{alpha2})--(\ref{beta2}) and we used (\ref{Dadef}) to express $\Dim,\,\Dik$ in terms of $m'_{qi}\leq 0$ and $k'_{qi}\leq 0$ (though $m_{qi}$ and $k_{qi}$ are necessarily non--negative). This condition (which is equivalent to the Hellaby--Lake condition (\ref{noshx2})) can be rephrased in terms of $\HH'_{qi},\,\hOmi'$ by means of the relation (\ref{OmD}) in Appendix E.
\begin{figure}[htbp]
\begin{center}
\includegraphics[width=4.5in]{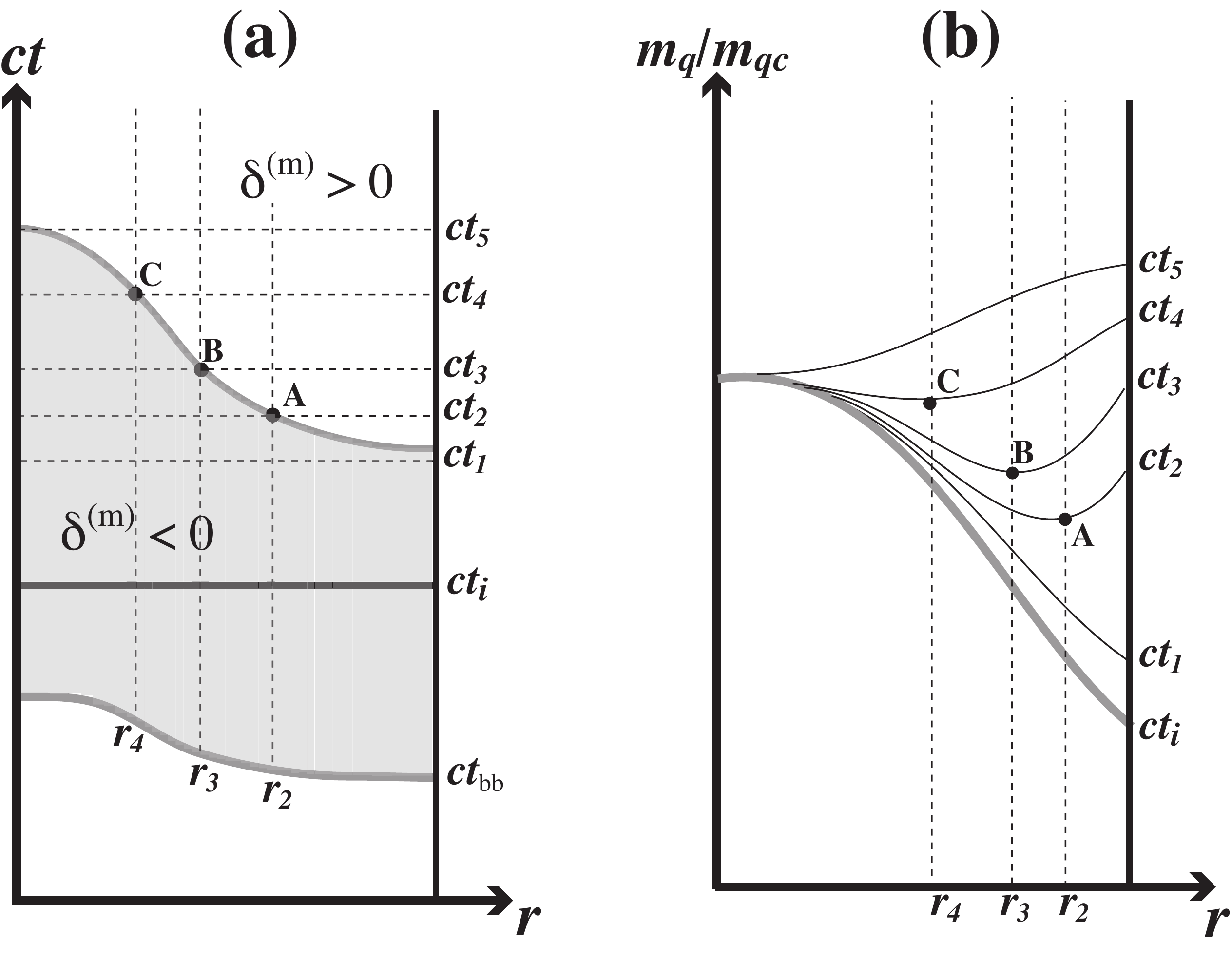}
\caption{{\bf Profile inversions with turning values.} Panel (a) displays the $(ct,r)$ plane associated with a radial domain (\ref{vartheta}). The region in which $\Dm(ct,r)<0$ is shaded, so that its upper boundary marks the surface $\Dm(ct,r)=0$ (upper thick curve). The locus of the initial singularity $c\tbb(r)$ is the lower thick curve. The black dots {\bf A},\,{\bf B},\,{\bf C} mark the points where $\Dm=0$ intersects $r=r_2,\,r_3,\,r_4$ where there is a turning value of $m_q$ for $ct=ct_2,\,ct_3,\,ct_4$. Panel (b) displays qualitative plots of the radial profiles of $m_q$ normalized to the central values $m_{qc}=m_q(ct,0)$ for the same values of $ct$ and $r$ shown in panel (a). Notice that an initial ``clump'' profile was selected, so that $\Dim\leq 0$ and $m'_{qi}\leq 0$ hold in all the radial domain, but for times $ct\geq ct_5$ this initial profile has evolved into a void profile with $\Dm\geq 0$ and $m'_q\geq 0$.  Notice (from Lemma 3) that turning values of $m_q$ imply turning values of its local dual scalar $m$. The situation described in this figure for $m_q$ is applicable to turning values and profile inversions of other scalars like $k_q$ or $\HH_q$ in hyperbolic models (as $\Dh$ diverges as $\HH_q\to 0$ in elliptic models, see section 8). Notice that the surface $\Dm(t,r)=0$ is the so--called ``density wave'' \cite{kras1}.}
\label{fig1}
\end{center}
\end{figure}
\begin{figure}[htbp]
\begin{center}
\includegraphics[width=4.5in]{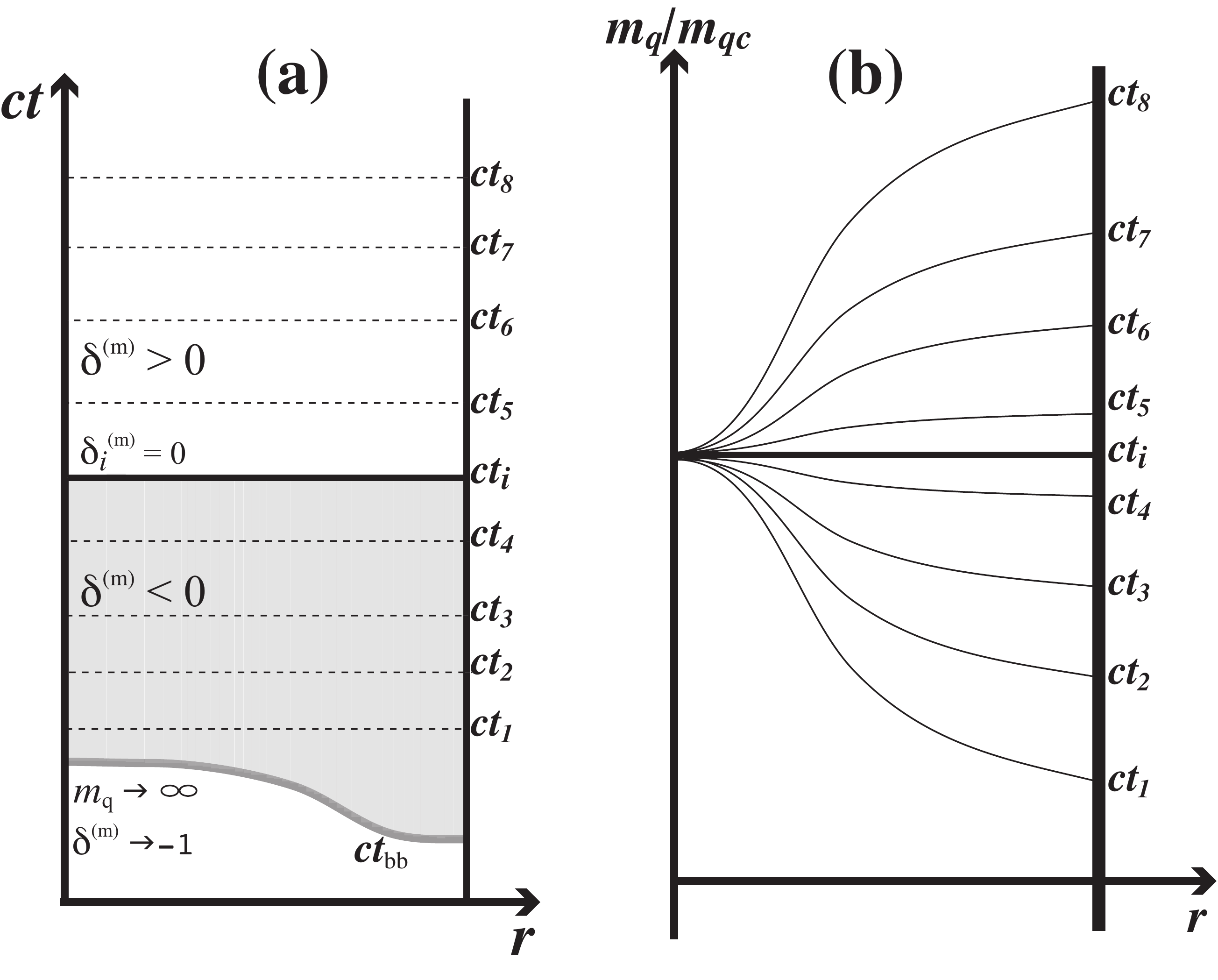}
\caption{{\bf Profile inversions without turning values.} Panel (a) displays the $(ct,r)$ plane associated with a radial domain (\ref{vartheta}). The region in which $\Dm(ct,r)<0$ is shaded, so that its upper boundary marks the simultaneous surface $\Dim(r)=0$, which corresponds to $ct=ct_i$ (upper thick line segment). The locus of the initial singularity $c\tbb(r)$ is the lower thick curve.  Panel (b) displays qualitative plots of the radial profiles of $m_q$ normalized to the central values $m_{qc}=m_q(ct,0)$ for the same values of $ct$ and $r$ shown in panel (a). Notice that $\Dim= 0$ implies $m'_{qi}= 0$ so that $m_{qi}=m_{qc}$, marking $ct=ct_i$ the locus of the profile inversion, as $m_q$ has clump profiles for $ct< ct_i$ and void profiles for $ct> ct_i$, without density TV's in any value of $t$. Notice (from Lemma 3) that turning values of $m_q$ imply turning values of its local dual scalar $m$. The same picture is applicable to profile inversions of $k_q$ or $\HH_q$.}
\label{fig2}
\end{center}
\end{figure}

\section{Profile inversions and turning values.}

In reference \cite{mushel} Mustapha and Hellaby addressed the problem of whether a density profile of the type of a clump(void) may be ``inverted'' into that of a void(clump) as a regular LTB model evolves in time. These authors examined the concavity of the radial density profile around the center of symmetry, but their results involve quite cumbersome second and third order radial derivatives of $R$ and the free functions $M,\,E,\,c\tbb$. Their proof of the existence of this inversion (their section 5) is based on restricting assumptions and is not fully general. In the remaining of this article we extend and complement these results and show by using Lemmas 2 and 3 that the issue of a ``profile inversion''  can be addressed rigorously and in a much more intuitive way, not only for density but also for spatial curvature and the expansion.

In order to address the ``clump into void'' question in precise and rigorous terms, we consider a quasi--local scalar function $A_q$ defined in a radial range (\ref{vartheta}) in a regular LTB model complying with condition (\ref{noshxG}), so that $R'>0$ ($\T[t]$ homemorphic to $\mathbb{R}^3$, see section 9 for the case of $\T[t]$ homeomorphic to $\mathbb{S}^3$).\\

\begin{quote}

\noindent \underline{Definition: Profile inversion}.  The scalar $A_q$ undergoes a profile inversion in a domain $\vartheta_c(r)$ if there exist two values $t=t_1,\,t_2$ with $t_1<t_2$, such that for all $x\in \vartheta_c(r)$ 
\begin{equation}\fl \hbox{sign}\left[\Da\right]\quad\hbox{for}\quad t<t_1\quad\ne\quad \hbox{sign}\left[\Da\right]\quad\hbox{for}\quad t>t_2, \label{prof_inv_def}\end{equation}
The ``clump into void'' inversion of $A_q$ corresponds to the particular case
\begin{equation} \fl \Da\leq 0\quad\hbox{for}\quad t<t_1\quad\hbox{and}\quad \Da\geq 0\quad\hbox{for}\quad t>t_2,\label{cintov_inv}\end{equation}
where the equal signs above hold only at the center. The opposite inversion (void into clump) is defined by the opposite signs.  

\end{quote}

\noindent
\noindent \underline{Comment}. It is useful to identify $t_1$ with $t_i$, so that the sign of the initial value function $\Da_i$ conveys the initial clump or void nature of the profile that becomes inverted for some $t_2>t_i$. However, $t_1 < t_i <t_2$ is also possible, which means selecting $A_{qi}$ whose profile is not monotonous.  

\begin{quote}

\noindent \underline{Lemma 4}. The existence of a profile inversion of $A_q$ implies a profile inversion of $A$.\\

\noindent \underline{Proof}. Following Lemma 1 and its corollary, a clump (void) of $A_q$ in a given domain $\vartheta_c(r)$ implies a clump (void) of $A$. Hence, the result follows. It is important to remark that for practical purposes it may be far easier to examine the profile inversion on $A_q$, so that the existence of a profile inversion of $A$ follows from Lemmas 1 and 3.\\

\noindent \underline{Lemma 5}. A sufficient (but not necessary) condition for a profile inversion of $A_q$ in a domain $\vartheta_c(r)$ is the existence of TV's marked by $x=\rtv[t]$ in every $\T[t]$ with $t\in\{t_1\leq t\leq t_2\}$ and with either one of $\rtv[t_1]=0$ or $\rtv[t_2]=0$ holding.\\ 

\noindent \underline{Proof}. Consider a clump into void inversion for $A_q>0$ and the case $\rtv[t_2]=0$ (the void into clump inversion and the cases $\rtv[t_1]=0$ or negative $A_q$ are analogous). Since there is no TV before $t_1$, then $A'_q< 0$ and $\Da\leq 0$ hold for all $x\in \vartheta_c(r)$ for $t<t_1$. For each $t_j\in\{t_1\leq t\leq t_2\}$ there is a TV at $x=\rtv[t_j]$ so that $\Da(ct_j,\rtv[j])=0$, and so we have:\, $A'_q< 0$ for $0<x<\rtv[t_j]$ and $A'_q> 0$ for $\rtv[t_j]<x<r$, so that $\rtv[t_j]\ne \rtv[t_k]$ for every $t_j\ne t_k$. As a consequence,  (\ref{lemma2}) is not marked by a surface of simultaneity in the $(ct,r)$ plane. Since there are no more TV's for $t>t_2$ in $\vartheta(r)$, then $A'_q> 0$ and $\Da\geq 0$ hold and the clump profile has been inverted. This process is illustrated by figure \ref{fig1}.\\

\noindent \underline{Corollary 1}. Lemma 5 also provides a sufficient condition for a profile inversion of $A$. This follows from Lemmas 2 and 3, as the sufficient condition for the emergence of a turning value ($A'=0$) at some $r=\rtv$ is simply (\ref{lemma2}).\\

\noindent \underline{Corollary 2}. If the initial value function $A_{qi}$ is selected so that $A'_{qi}=\Da_i=0$ for $r=\rtv[t_i]\in \vartheta_c(r)$ and there exist $\bar t>t_i$ such that $\Da\ne 0$ for all $t>\bar t$ and $r>0$, then there is a profile inversion of $A_q$ in the domain $\vartheta_c(\rtv[t_i])\subset \vartheta_c(r)$. From Lemmas 3 and 4, there will also be a profile inversion of $A$. This is a restricted form of Lemma 5 in which there is no information on the sign of $\Da$ for $t<t_i$.\\

\noindent \underline{Comment}. The choice of an initial value function $A_{qi}$ with a non--monotonous profile (there exists a TV such that $\Da_i=0$ holds for some $r=\rtv[t_i]\in \vartheta_c(r)$) is not sufficient for having a profile inversion, since $\Da$ may keep the same sign for all $t>t_i$ in the domain  $\vartheta_c(\rtv[t_i])$. In general, further information on $\Da$ is needed given a non--monotonous profile of $A_{qi}$.

\end{quote}

\noindent
We have provided sufficient conditions for a profile inversion of $A$ based on the existence of an inversion of $A_q$ and TV's, however, such inversions can also occur without TV's. 

\begin{quote}

\noindent \underline{Lemma 6}. If there are no TV's of $A_q$, a sufficient (not necessary) condition for a profile inversion at a fixed $t=t_0$ follows if (\ref{lemma2}) is marked by a surface of simultaneity $t=t_0$
\begin{equation}\Da(ct_0,x)=0\quad \forall \quad x\in\vartheta_c(r),\label{prof_inv_noTV}\end{equation}
so that $\Da$ has opposite signs for $t<t_0$ and for $t>t_0$. \\

\noindent \underline{Proof}. Consider the case of a clump into void inversion and $A_q>0$. Since there are no TV's, then $\Da<0$ in $\vartheta_c(r)$ for all $t<t_0$ and $\Da>0$ in $\vartheta_c(r)$ for all $t>t_0$. From Lemmas 1--3, the sign of $\Da$ determines the sign of $A'_q$, hence (since there are no TV's) a profile inversion occurs at $t=t_0$. This situation is illustrated by figure \ref{fig2}. Notice that $\Da=0$ at $t=t_0$ does not imply that $\Da$ vanishes at $t\ne t_0$. 

\end{quote}

\noindent
Given the definition of a a profile inversion and its relation with the existence of TV's in the radial profile of scalars, we need to examine now the conditions for $\Da=0$ for the main quasi--local scalars $A_q=m_q,\,k_q,\,\HH_q$. As we show in the following sections these conditions are easier to analyze and more intuitive that the concavity test around the center considered in \cite{mushel}.

\section{Radial density profiles and their inversions.}

Following Lemma 1, the qualitative clump or void form of the radial profile of $m_q=(\kappa/3)\rho_q$ is determined by the sign of $\Dm$. From Lemmas 1 and 2 this information readily translates into general statements on the profile of local density $m=(\kappa/3)\rho$. To examine the sign of $\Dm$ we obtain this function directly from (\ref{slawDm}), which by using (\ref{Ghe}) and (\ref{Ghe}) leads to
\bse\ba \fl \Dm = \frac{\Dim}{\Gamma L^{3/2}}=\frac{\Dim}{(1+\Dim)\,L^{3/2}-\Dim},\quad\hbox{parabolic}\label{Dmposp}\\
     \fl \Dm = \frac{3}{\Gamma}\left\{\left(\Dim-\frac{3}{2}\Dik\right)\left[\HH_q c(t-t_i)-\frac{2}{3}\right]+\left(\Dim-\Dik\right)\frac{\HH_q}{\HH_{qi}}\right\},\nonumber\\\fl\hskip 8cm\hbox{hyperbolic and elliptic}\label{Dmposhe}\ea\ese
where $\Gamma,\, \HH_q$ and $\HH_{qi}$ are given by (\ref{Hq}), (\ref{Gp}), (\ref{Ghe}) and (\ref{Hqqi}), and we assume that initial value functions $\Dim,\,\Dik$ and $\HH_{qi}$ have been selected so that the Hellaby--Lake conditions  (\ref{noshxGh}) and (\ref{noshxGe}) hold ($\Gamma>0$).

Substituting $\Gamma$ from (\ref{Gp}) and (\ref{Ghe}) into (\ref{Dmposp}) and  (\ref{Dmposhe}), and bearing in mind that $\HH_q\to \infty$ as $t\to\tbb$ (or $L\to 0$, see Appendix B), we find that in this limit
\begin{equation} \Dm\to-1,\label{Dmtbb}\end{equation}
holds for parabolic, hyperbolic and elliptic models and for whatever initial value functions that we might choose at $t=t_i$. The same limiting value holds as $t\to\tcoll$ in the collapsing singularity in elliptic models (where $\HH_q\to-\infty$ as $L\to 0$, see Appendix B). From (\ref{slawDm}), the limiting behavior (\ref{Dmtbb}) necessarily implies that $\Gamma\to\infty$ as $L\to 0$, as in general $\Dim>-1$. We also assume domains $\vartheta_c(r)$ given by (\ref{vartheta}) in which $k_{qi}$ is either zero or does not change sign in the full domain,  which means that $\vartheta_c(r)$ defines a ``pure'' parabolic, hyperbolic or elliptic region containing a center (possibly part of a ``pure'' model) in which $\Dik$ is bounded (mixed elliptic/hyperbolic configurations are examined in section 10).   

\subsection{Profile inversion: ``Density clumps into voids'' revisited.}  

A density profile inversion of $m_q$ in a regular LTB model complying with (\ref{noshxG}) is characterized by (\ref{cintov_inv}) for $A=m_q$. From Lemma 4, (\ref{cintov_inv}) also implies a profile inversion for the local density $m$. Considering domains of the form $\vartheta_c(r)$, sufficient conditions for the profile inversion of $m_q$ can arise if 
\begin{itemize}
\item There exist density turning values ($m'_q=0$ and $\Dm=0$) at $r=\rtv[t]$ for slices $\T[t]$ marked by $t_1\leq t\leq t_2$ with $t_i<t_1$ (Lemma 5, figure \ref{fig1}).
\item If there are no TV's, then there must exist $t=t_0$ such that $\Dm(ct_0,x)=0$ and the sign of $\Dm$ for $t>t_0$ is the opposite to that for $t<t_0$ (Lemma 6, figure \ref{fig1}).
\end{itemize}
We examine the conditions for the fulfillment of these two possibilities.     

Consider the case of regular parabolic models. It is evident from (\ref{noshx_par}) and (\ref{Dmposp}) that $\Dm\leq 0$ must hold in any radial domain $\vartheta_c(r)$ in these models (with the equal sign holding only at the center). Hence a density profile inversion is not possible: their density profiles are monotonously decreasing for all $t$ so that initial parabolic clumps remain clumps for all the time evolution.  For domains in hyperbolic or elliptic configurations the Hellaby--Lake conditions (\ref{noshxGh}) and (\ref{noshxGe}) do not exclude {\it {a priori}} the possible emergence of a density profile inversion, but checking if conditions (\ref{cintov_inv}) or (\ref{prof_inv_noTV}) can be fulfilled requires further examination. In practice, we need to examine how $\Dm$ in (\ref{Dmposhe}) evolves in time given the existence (or lack) of turning values. This process is illustrated by figures \ref{fig1} and \ref{fig2}.

\subsection{Hyperbolic models or regions.}

\subsubsection{Inversions with TV's}

We assume henceforth that initial value functions have been selected so that (\ref{noshxG}) and thus, the Hellaby--Lake conditions (\ref{noshxGh}), hold for all $t$. For arbitrary fixed (and finite) $r$ and large times $t/t_i\gg 1$  we have from (\ref{Hq}):\, $\dot L\approx |k_{qi}|^{1/2}$, which leads to $L\approx |k_{qi}|^{1/2}(t/t_i)\gg 1$. Hence the large time asymptotic regime follows from asymptotic series for $L\gg 1$:
\bse\ba \frac{\HH_q}{\HH_{qi}}\approx \frac{x_i^{1/2}}{(2+x_i)^{1/2}\,L}+O(L^{-2}),\label{asHHi}\\
\HH_q\,c(t-t_i)\approx 1-\frac{2}{x_i}\,\frac{\ln L}{L}+O(L^{-1}),\label{asHct}\ea\ese
where $c(t-t_i)$ follows from (\ref{hypZ2}). These asymptotic forms imply that $\HH_q/\HH_{qi}$ and $\HH_q c(t-t_i)$ in (\ref{Ghe}) and (\ref{Dmposhe}) respectively decrease and increase monotonically between $0$ and $1$. The asymptotic for of $\Dm$ for $t/t_i\gg 1$ is 
\begin{equation} \Dm\approx \frac{\Dim-(3/2)\Dik}{1+(3/2)\Dik}\,\left[1-\frac{2}{x_i}\,\frac{\ln L}{L}+O(L^{-1})\right],\label{asDm} \end{equation}
where we used (\ref{Ghe}),  (\ref{Dmposhe}) and (\ref{asHHi})--(\ref{asHct}). Therefore, we have the following general result:\\

\begin{quote}

\noindent \underline{Lemma 7}. If $\Dim\ne 0,\,\Dik\ne 0$ comply with the Hellaby--Lake conditions (\ref{noshxGh}), a necessary and sufficient condition for the existence of a ``clump to void'' density profile inversion in a regular hyperbolic model is
\begin{equation}  \Dim-\frac{3}{2}\Dik \geq 0,\label{dm32dk}\end{equation}
where the equal sign holds only at the center.\\ 

\noindent \underline{Proof}.  In the asymptotic regime $t/t_i\gg1 $ the form of the leading term of $\Dm$ in (\ref{asDm}) becomes independent of $x_i$ (or $m_{qi},\,k_{qi}$) for sufficiently large $L$, hence (\ref{dm32dk}) implies $\Dm\geq 0$ for all $r$ in this limit and the converse statement is also true. On the other hand, $\Dm\to -1$ as $L\to 0$ for all $x\in\vartheta_c(r)$ is necessary and sufficient if the Hellaby--Lake conditions to hold. Hence, $\Dm$ must change sign at least once if (\ref{dm32dk}) holds and the converse statement is true (see figure \ref{fig1} for a graphic depiction).\\

\noindent \underline{Corollary}. A ``void to clump'' inversion without shell crossings is not possible. 

\smallskip 
\noindent
{\underline{Proof}}. A ``void into clump'' scenario requires the combination $\Dim\geq 0$ and $\Dm\leq 0$, which would only have a chance to occur if $\Dik\geq 0$, but this initial condition together with $\Dim\geq 0$ violates (\ref{tbbrh}), and so is incompatible with $\Gamma>0$ (and consequently, with the Hellaby--Lake conditions).\\  

\noindent \underline{Comment: no profile inversion}. The converse statement of Lemma 7 implies that the reverse condition, $\Dim-(3/2)\Dik\leq 0$ (plus (\ref{noshxGh})), indicates an shell--crossing free evolution without a density profile inversion: the initial density clump profile is preserved (see Table 1). Notice that a void profile for the whole time evolution is not possible because near the initial singularity (as $t\to\tbb$) we have necessarily $\Dm\to -1$ for every choice of initial conditions. However, the pure void evolution is possible for models with a simultaneous big bang ($\tbb'=0$) for which $\Dm\to 0$ as $t\to\tbb$ (see section 10).  

\end{quote}

\noindent
We examine now the consequences of Lemma 7 for different assumptions on $\Dim$ and $\Dik$. 
If we choose an initial density with a clump profile ($-1<\Dim\leq 0$ so that $\Dim=-|\Dim|$) and desire an evolution with a clump to void evolution, then the profile inversion must occur for times $t>t_i$, irrespectively of the existence of density TV's. In this case, condition (\ref{dm32dk}) implies  
\begin{equation} \Dik\leq 0,\qquad |\Dik|>\frac{2}{3}|\Dim|.\label{CintoV1}\end{equation}
If we assume an initial density void $\Dim\geq 0$, together with $\Gamma>0$, then a profile inversion must have already happened for previous times  $\tbb <t<t_i$ because $\Dm\to -1$ as $t\to\tbb$ (with ot without density TV's). Also, if $\Dim\geq 0$ and $\Gamma>0$ hold, then  (\ref{asDm}) implies an asymptotic void profile $\Dm\geq 0$ if $\Dik\leq 0$. The allowed qualitative evolution of the radial profiles of density and spatial curvature are depicted in Table 1 for all possible sign combinations of $\Dim$ and $\Dik$ compatible with (\ref{noshxG}). 

\subsubsection{Inversions without TV's}

While Lemma 7 applies to either one of the possible situations described by Lemmas 5 and 6, the clump into void inversion without density TV's is compatible with less restrictive forms of the initial value functions. Since $t=t_0$ is arbitrary, is is convenient to examine this case by assuming that $t_0=t_i$, leading to

\begin{quote}

\noindent \underline{Lemma 8}. If there are no density TV's along the hypersurfaces $\T[t]$ of a regular hyperbolic model, a sufficient condition for a clump to void density profile inversion in $\vartheta_c(r)$ is
\begin{equation} \Dim = 0,\qquad -\frac{2}{3}<\Dik\leq 0.\label{CintoV2}\end{equation}
\noindent \underline{Proof}. If $\Dim=0$, then $\Dm$ in (\ref{Dmposhe}) takes the form
\begin{equation}\fl \Dm = \frac{3\Dik}{\Gamma}\,\left[1-\frac{\HH_q}{\HH_{qi}}-\frac{3}{2}\,\HH_q\,c(t-t_i)\right]\approx -\frac{(3/2)\,\Dik}{1+(3/2)\,\Dik},\label{CintoV3}\end{equation}
where we substituted $\Dim=0$ in the expression for $\Gamma$ in (\ref{Ghe}) and used only the leading term in the asymptotic expansions (\ref{asHHi})--(\ref{asHct}). Evidently, $\Dm$ in (\ref{CintoV3}) implies the conditions on $\Dik$ specified in (\ref{CintoV2}). Since $\Dim=0$ and $\Dm>0$ as $t\to\infty$, and there are no density TV's, then $\Dm>0$ holds for all $t>t_i$. On the other hand, since $\Dm\to -1$ as $t\to\tbb$ (or $L\to 0$), then the lack of density TV's implies that $\Dm<0$ for all $t<t_i$. Hence, a clump to void (and not a void to clump) density profile inversion must occur at $t=t_i$ (see figure \ref{fig2}).   

\end{quote}

\subsubsection{Initial conditions that yield inversions.}

\noindent
It is important to remark that initial conditions complying with a clump into void density profile inversion are not strange nor unusual, and are perfectly compatible with the Hellaby--Lake regularity conditions (\ref{noshxGh}). Condition (\ref{dm32dk}) can be expressed in terms of gradients of initial value functions by the equivalent inequalities:
\bse\ba  \frac{m'_{qi}}{m_{qi}}-\frac{3}{2}\frac{k'_{qi}}{k_{qi}}\geq 0, \quad\hbox{with:}\quad m'_{qi}\leq 0,\,\, k'_{qi}\geq 0,\,\, k_{qi}\leq 0,\label{CVmk1}\\
\frac{M'}{M}-\frac{3}{2}\frac{E'}{E}\geq 0,\quad\hbox{with:}\quad 0<\frac{M'}{M}\leq \frac{3R'_i}{R_i},\quad0<\frac{E'}{E}\leq \frac{2R'_i}{R_i}, \label{CVME1}\ea\ese
while condition (\ref{CintoV2}) is equivalent to
\bse\ba m'_{qi} = 0,\,\, k'_{qi}\geq 0,\,\, k_{qi}\leq 0,\label{CVmk2}\\ 
\frac{M'}{M} = \frac{3R'_i}{R_i},\quad0<\frac{E'}{E}\leq \frac{2R'_i}{R_i}, \label{CVME2}\ea\ese
Notice that these conditions are compatible with the constraints for an initial clump in regular models given in (\ref{clumpME}) and (\ref{voidME}). Also, we can rewrite (\ref{CVmk1})--(\ref{CVME1}) and (\ref{CVmk2})--(\ref{CVME2}) in terms of the radial coordinate by means of the coordinate gauge (\ref{rgauge}).

\subsection{Elliptic models or regions.}

The following general result holds for elliptic models or regions in which $R'>0$ holds everywhere (no TV of $R$):

\begin{quote}

\noindent \underline{Lemma 9}. There are no density profile inversions (with or without density TV's) in any domain $\vartheta_c(r)$ in a regular elliptic model or region lacking a TV of $R$. The proof of the is given below.\\

\noindent \underline{Corollary}. Density has a clump profile for all the evolution time and for every domain $\vartheta_c(r)$, hence $-1<\Dm\leq 0$ (clump profile) must hold everywhere. If Lemma 9 holds, the proof follows directly from (\ref{slawDm}) and (\ref{noshxGe}).  

\end{quote} 

\noindent
{\underline{Proof of Lemma 9}}. We examine first the conditions for a profile inversion when there are density TV's. Regularity implies that $\Gamma>0$ holds for all times, so that (\ref{CVell1a})--(\ref{CVell1c}) hold. Hence, in order to examine the conditions for a density profile inversion in these configurations, it is useful to rewrite conditions (\ref{noshx_nec1}), (\ref{noshxce}) and (\ref{noshx_nec2}), which are necessary and necessary and sufficient for $\Gamma>0$, as
\bse\ba \Dim=-|\Dim|,\qquad \Dik=-|\Dik|,\label{CVell1a}\\
\Dim-\frac{3}{2}\Dik\geq 0\quad \Rightarrow\quad |\Dik|\geq \frac{2}{3}|\Dim|,\label{CVell1b}\\
|\Dik| \leq \frac{\alpha_1}{\beta_1}|\Dim|\leq \frac{\alpha_1\beta_2}{\beta_1\alpha_2}|\Dik|,\quad\hbox{with:}\quad \frac{\alpha_1\beta_2}{\beta_1\alpha_2}>\frac{\alpha_1}{\beta_1}>1.\label{CVell1c}\ea\ese
where $\alpha_1,\,\alpha_2,\,\beta_1,\,\beta_2$ are given by (\ref{alpha1})--(\ref{beta1}) and (\ref{alpha2})--(\ref{beta2}). Considering (\ref{CVell1a}), we rewrite (\ref{Dmposhe}) as
\begin{equation}\fl\Dm = \frac{-|\Dim|+3C}{1-3C},\qquad C\equiv |\Dim|\Phi-|\Dik|\Psi<\frac{1}{3},\label{DmEll}\end{equation}
where
\bse\ba \Phi \equiv 1-\frac{\HH_q}{\HH_{qi}}-\HH_q c(t-t_i),\label{Phi}\\
\Psi \equiv 1-\frac{\HH_q}{\HH_{qi}}-\frac{3}{2}\HH_q c(t-t_i),\label{Psi} \ea\ese
From the expressions above we can readily examine the behavior of $\Dm$ at maximal expansion, $t=\tmax$ (or $L=\Lmax=2/x_i$) where $\HH_q= 0$, as well as near the collapse $t\to\tcoll$ where $\HH_q\to-\infty$. The result is
\bse\ba\Dm= -3\frac{|\Dik|- (2/3)|\Dim|}{1-3|\Dim|+3|\Dik|},\qquad t=\tmax\label{Dmtmax}\\\
\Dm\to -1,\qquad t\to\tcoll.\label{Dmtcoll}\ea\ese
Comparing (\ref{Dmtmax}) with the necessary condition (\ref{CVell1b}), it is evident that $\Gamma>0$ implies that $\Dm\leq 0$ at $t=\tmax$. Since $\Dm\leq 0$ at $t=\tmax$ and as $t\to\tcoll$, the only possibility for $\Dm\geq  0$ to occur is for some intermediate times $t_i<t<\tmax$ and/or $\tmax<t<\tcoll$.  We now assume that $\Dm>0$ and $\Gamma>0$ both hold for these times and verify if this assumption is consistent with (\ref{CVell1a})--(\ref{CVell1c}).

If $\Dm>0$ and $\Gamma>0$ both hold we have from (\ref{DmEll})
\begin{equation}  |\Dik|<\frac{3\Phi-1}{3\Psi}\,|\Dim|< |\Dik|+\frac{1-|\Dim|}{3\Psi},\label{Dmposcond}\end{equation}
an inequality that can be examined qualitatively by means of the analytic expressions (\ref{ellZ2}) and (\ref{ellZ3}). In the range $t_i<t<\tmax$ the term $(3\Phi-1)/(3\Psi)$ grows monotonically from $-\infty$ to $2/3$, hence (\ref{Dmposcond}) implies that $|\Dik| <(2/3)|\Dim|$ holds in all this range, which is in contradiction with the necessary condition (\ref{CVell1b}) and thus implies that $\Dm>0$ cannot hold. 

Now consider the range $\tmax<t<\tcoll$, in which the term $(3\Phi-1)/(3\Psi)$ grows monotonically from $2/3$ to a maximal value as $L\to 0$ (or $t\to\tcoll$) given by
\begin{equation} \frac{3\Phi(0,x_i)-1}{3\Psi(0,x_i)}= \frac{\alpha_2(x_i)}{\beta_2(x_i)}<1,\qquad x_i=\frac{k_{qi}}{m_{qi}}=\frac{2(\hOmi-1)}{\hOmi}.\label{maxval}\end{equation}
where $0<x_i\leq 2$ and $\alpha_2,\,\beta_2$ are given by (\ref{alpha2})--(\ref{beta2}), while $\hOmi$ is defined in Appendix E. Since now (\ref{Dmposcond}) is not in contradiction with (\ref{CVell1b}) because it allows for $|\Dik|>2/3|\Dim|$, we need to see if it is compatible with (\ref{CVell1c}). Combining the latter and (\ref{Dmposcond}) leads to the inequality
\begin{equation} \frac{\alpha_2(x_i)}{\beta_2(x_i)}|\Dim|<|\Dik|<\frac{3\Phi(L,x_i)-1}{3\Psi(L,x_i)}\,|\Dim|,\label{Dmposcond2}\end{equation}
which should hold for all $x_i$ and $L$. Since (\ref{maxval}) is valid for all initial value functions, then (\ref{Dmposcond2}) is clearly inconsistent, and so (\ref{Dmposcond}) is inconsistent with (\ref{CVell1c}) and so $\Dm>0$ cannot hold in this range of $t$.

For a profile inversion without TV's we need to impose the condition $\Dim=0$, as in (\ref{CintoV2}) (see Lemma 8). From (\ref{tmc1}) and (\ref{tbbr}) and (\ref{tcollr}) two of the Hellaby--Lake conditions in (\ref{noshxGe}) now take the form
\bse\ba\frac{c\tbb'}{3R'_i/R_i}=\frac{\Dik}{\HH_{qi}}\left[\frac{3Z_e(x_i)(2-x_i)^{1/2}}{2x_i^{3/2}}-1\right]\leq 0,\label{HL11}\\
\frac{c\tcoll'}{3R'_i/R_i} = -\frac{3}{2}\Dik c(\tcoll-\tbb)+\frac{c\tbb'}{3R'_i/R_i}\geq 0.,\label{HL12}\ea\ese
Since the term in square brackets in (\ref{HL11}) is negative for all the allowed range $0<x_i\leq 2$, fulfillment of this regularity condition requires $\Dik\geq 0$, but then condition (\ref{HL12}) will not be satisfied. Therefore, there are no regular elliptic models complying with the conditions for a profile inversion without TV's.

\begin{table}
\begin{center}
\begin{tabular}{|c| c| c| c| c| c| c|}  
\hline
\hline
{Case} &{$\Dim$} &{$\Dik$} &{$\Dim-\frac{3}{2}\Dik$} &{$\Dm$} &{$\Dk$} &{Comments}
\\ 
\hline
\hline
{} &{} &{} &{} &{} &{} &{No profile inversion of $k$.}
\\
{i} &{$\leq 0$} &{$\leq 0$} &{$\geq 0$} &{$\geq 0$}  &{$\leq 0$} &{Profile inversion }
\\
{} &{} &{} &{} &{} &{} &{of $m$ at $t>t_i$}
\\
\hline
\hline
{} &{} &{} &{} &{} &{} &{No profile inversion of $m$.}
\\
{ii} &{$\leq 0$} &{$\leq 0$} &{$\leq 0$} &{$\leq 0$}  &{$\geq 0$} &{Profile inversion}
\\
{} &{} &{} &{} &{} &{} &{of $k$ at $t>t_i$}
\\
\hline
\hline
{} &{} &{} &{} &{} &{} &{No profile inversion of $m$.}
\\
{iii} &{$\leq 0$} &{$\geq 0$} &{$\leq 0$ (nec)} &{$\leq 0$}  &{$\geq 0$} &{Profile inversion}
\\
{} &{} &{} &{} &{} &{} &{of $k$ at $t<t_i$}
\\
\hline
\hline
{} &{} &{} &{} &{} &{} &{No profile inversion of $k$.}
\\
{iv} &{$\geq 0$} &{$\leq 0$} &{$\geq 0$ (nec)} &{$\geq 0$}  &{$\leq 0$} &{Profile inversion}
\\
{} &{} &{} &{} &{} &{} &{of $m$ at $t<t_i$}
\\
\hline
\hline
\end{tabular}
\end{center}
\caption{{\bf{Initial conditions and radial profiles of density and spatial curvature in hyperbolic models}}. The table presents a summary of the qualitative evolution of the radial profiles of $m$ and $k$ for all possible combinations of $\Dim\ne 0$ and $\Dik\ne 0$, compatible with regularity conditions (\ref{noshxGh}). The fifth and sixth columns display the asymptotic sign of $\Dm$ and $\Dk$. The term ``nec'' means that $\Dim-(3/2)\Dik$ necessarily has the indicated sign for these specific sign choices of $\Dim$ and $\Dik$.}
\end{table}


\section{Radial profiles of spatial curvature.}

Just as with the density, the qualitative behavior of radial profiles of $k_q=\RR_q/6$ and $k=\RR/6$ are determined by the sign of $\Dk$, which takes the following form for both hyperbolic and elliptic models (with $\T[t]\sim\mathbb{R}^3$):
\begin{equation} \fl \Dk = \frac{2}{\Gamma}\left\{\left(\Dim-\frac{3}{2}\Dik\right)\left[\HH_q c(t-t_i)-1\right]+\left(\Dim-\Dik\right)\frac{\HH_q}{\HH_{qi}}\right\},\label{Dkposhe}\end{equation}
where $\Gamma,\,\HH_q$ and $c(t-t_i)$ are given by (\ref{Hq}), (\ref{Gp}), (\ref{Ghe}), (\ref{hypZ2}) and  (\ref{ellZ2}), and fulfillment of regularity conditions  (\ref{noshxGh}) and (\ref{noshxGe}) is assumed. Notice that $\Dk=0$ holds for all domains $\vartheta_c(r)$ in parabolic models and regions. 

As in the previous section, we now substitute $\Gamma$ in (\ref{Ghe}) into   (\ref{Dkposhe}). Bearing in mind that $\HH_q\to \infty$ as $t\to\tbb$ (or $L\to 0$), we find that in this limit
\begin{equation} \Dk\to-\frac{2}{3},\label{Dktbb}\end{equation}
holds for hyperbolic and elliptic models and for whatever initial value functions that we might choose at $t=t_i$. The same limiting value holds as $t\to\tcoll$ in the collapsing singularity in elliptic models (where $\HH_q\to-\infty$ as $L\to 0$). From (\ref{slawDk}), the limiting behavior (\ref{Dktbb}) necessarily implies that $\Gamma\to\infty$ as $L\to 0$. 

The profile inversion of spatial curvature was not considered by Mustapha and Hellaby, but this issue can also be examined along following the same arguments of the previous section. Table 1 summarizes the possible qualitative evolution of $k$ and $k_q$ for all combinations of initial value functions compatible with the regularity conditions (\ref{noshxGh}) and (\ref{noshxGe}).\\ 

\noindent
{\underline{Hyperbolic models or regions.}}\\

Bearing in mind the asymptotic expansions (\ref{asHHi})--(\ref{asHct}) and $\Gamma$ given in (\ref{Ghe}), the form of $\Dk$ in (\ref{Dkposhe}) for $L\gg 1$ is
\begin{equation} \Dk\approx -2\,\frac{\Dim-(3/2)\Dik}{1+(3/2)\Dik}\,\left[\frac{2}{x_i}\,\frac{\ln L}{L}+O(L^{-1})\right].\label{asDk} \end{equation}
Hence, $\Dk\to 0$ as $L\to\infty$ (or $t\to\infty$), but its sign is determined by the opposite condition to the sign of $\Dm$, which means that the features of the radial profile evolution, including conditions for a profile inversion of $k$, are given by the complementary situations to those of $m$. As a consequence, the sign combinations of $\Dim,\,\Dik$ and $\Dim-(3/2)\Dik$ that yield a given sign for $\Dm$, will yield the opposite sign for $\Dk$. \\ 

\noindent
{\underline{Elliptic models or regions.}}\\

As in the previous section, we rewrite conditions (\ref{noshx_nec1}), (\ref{noshxce}) and (\ref{noshx_nec2}) as in (\ref{CVell1a})--(\ref{CVell1c}), transforming (\ref{Dkposhe}) into the following form similar to (\ref{DmEll})
\begin{equation}\Dk = \frac{-|\Dik|+2C}{1-3C},\label{DkEll}\end{equation}
where $C$ is given by (\ref{DmEll}) and (\ref{Phi})--(\ref{Psi}). Since $\Dk\to -2/3$ as $t\to\tbb$ and $t\to\tcoll$, whereas $\Dk$ tends to the same limit (\ref{Dmtmax}) as $t\to\tmax$, then following the same steps as in case of $\Dm$ in the previous section we can prove that $\Dk\leq 0$ holds for all the evolution range $\tbb<t<\tcoll$. Therefore, just as with density profiles, spatial curvature profiles in regular elliptic models whose hypersurfaces $\T[t]$ have $\mathbb{R}^3$ topology have the form of clumps for all $t$.

\section{Profiles of the expansion scalar}

From (\ref{Hq}), (\ref{qltransfH}), the scaling law (\ref{slawDh}) and (\ref{Omparab})--(\ref{Omell}) we can readily infer the sign of $\Dh$ from what we have found in previous sections about the signs of $\Dm$ and $\Dk$. For parabolic models we have $2\Dh=\Dm$, hence regularity conditions require $\Dh\leq 0$ ({\it{i.e.}} a clump radial profile for $\HH=\Theta/3$ and $\HH_q=\Theta_q/3$) for all the evolution time $t>\tbb$. 

For both hyperbolic and elliptic models or regions we have $2m_q/\HH_q^2=\hOm\to 1$ as $L\to 0$ ({\it{i.e.}} as $t\to\tbb$ and $t\to\tcoll$ in elliptic models). Hence, in this limit we have $\Dh\approx\Dm/2$, and so
\begin{equation} \Dh \to -\frac{1}{2}.\label{Dhtbb}\end{equation}
We examine now the sign of $\Dh$ for domains $\vartheta_c(r)$ in hyperbolic and elliptic configurations.\\

\noindent
{\underline{Hyperbolic models or regions.}}\\

From (\ref{Omhyp}) and (\ref{slawDh}), we have $0<\hOm<1$ and $\hOm\to 0$ for $L\gg 1$ (or $t\to\infty$). Hence $\Dh\to \Dk/2$ in this limit, and so $\Dh$ has the same asymptotic sign as $\Dk$ (see Table 1). Bearing in mind (\ref{Dhtbb}), there must be a profile inversion of $\HH$ and $\HH_q$ for the cases (ii) and (iii) in Table 1 for which $\Dim-(3/2)\Dik\leq 0$ and $\Dk\geq 0$ holds asymptotically:
\bse\ba \fl \hbox{Case ii}\qquad \Dim\leq 0,\,\Dik\leq 0,\qquad |\Dik|\leq \frac{2}{3}|\Dim|,\label{Hcase2}\\ \fl \hbox{Case iii}\qquad \Dim\leq 0,\,\Dik\geq 0,\label{Hcase3}\ea\ese
From (\ref{slawDh}), we have $\Dih\leq 0$ in case (ii) where both $\Dim$ and $\Dik$ are non--positive, hence in this case the profile inversion must occur for some $t>t_i$.  However, for case (iii) it is not evident if this inversion happens before or after $t=t_i$. We examine this case below.

The sign conditions of $\Dih$ for the case (iii) in (\ref{Hcase3}) can be given by
\bse\ba \fl \Dih \leq 0,\qquad \frac{\Dik}{|\Dim|}\leq \frac{2}{x_i},\quad \hbox{profile inversion at}\quad t>t_i,\label{Hcase31}
\\ \fl \Dih \geq 0,\qquad \frac{\Dik}{|\Dim|}\geq \frac{2}{x_i},\quad \hbox{profile inversion at}\quad t<t_i,\label{Hcase32}\ea\ese
where $|\Dim|\leq 1$ (because $\Dim\leq 0$). We need to verify the compatibility between (\ref{Hcase31})--(\ref{Hcase32}) and the regularity condition (\ref{tbbrh}), which for the sign combination (\ref{Hcase3}) takes the form
\begin{equation} \frac{\Dik}{|\Dim|}\leq \frac{\alpha(x_i)}{\beta(x_i)},\end{equation}
where $\alpha(x_i)$ and $\beta(x_i)$ are given by (\ref{alphah}) and (\ref{betah}). By looking at these expressions, we can readily see that $2/x_i <\alpha(x_i)/\beta(x_i)$ holds for the full range of $x_i$, which can be expressed as $0<\hOmi<1$ where $\hOmi=2/(2+x_i)$ (see Appendix E). Hence, all initial conditions (\ref{Hcase31}) associated  with $\Dih\leq 0$ will be compatible with (\ref{tbbrh}), whereas initial conditions (\ref{Hcase32}) associated with $\Dih\geq 0$ need to satisfy the extra constraint
\begin{equation} \frac{2}{x_i}\leq \frac{\Dik}{|\Dim|}\leq \frac{\alpha(x_i)}{\beta(x_i)}.\label{Hcase33} \end{equation}
As shown by these expressions, the sign of $\Dih$ (and thus the time of the profile inversion of $\HH$) is not obvious for the case (iii), and so it must be found by testing the fulfillment of (\ref{Hcase31})--(\ref{Hcase32}) and (\ref{Hcase33}) for any given choice of initial value functions.

A regular evolution without profile inversion of $\HH$ is also possible. The necessary (not sufficient) condition for this is given by cases (i) and (iv) in Table 1. However, it is not possible to provide a condition for this evolution without going to particular cases or specific restrictions of the initial value functions.\\  

\noindent
{\underline{Elliptic models or regions.}}\\

Near the initial and collapsing singularities ($\tbb$ and $\tcoll$) we have $\Dh\to -1/2$, but since $\HH_q\to 0$ and  $\hOm\to\infty$ as $t\to\tmax$, then $\Dh$ diverges in this limit. We examine then the radial profiles of the expansion by looking at the sign of 
\begin{equation}\HH_q\Dh = \HH-\HH_q = \frac{\HH'_q}{3\Gamma R'_i/R_i} = \frac{2m_q\,\Dm-k_q\,\Dk}{2\HH_q}.\end{equation}
where we used (\ref{Dadef}), (\ref{Omdef1})--(\ref{Omdef2}) and (\ref{Hqqi}). Notice, from (\ref{qltransfSE}), that this quantity is the negative of the scalar $\Sigma$ associated with the shear tensor (see \ref{SigEE1})).  From (\ref{slawDh}), (\ref{Ghe}), (\ref{Dmposhe}) and (\ref{Dkposhe}), it is straightforward to show that the leading term of $\HH_q\Dh$ for $L\approx L_{\textrm{\tiny{max}}}$ is
\ba\HH_q\Dh \approx \frac{m_{qi}}{L_{\textrm{\tiny{max}}}^3\left[1+3(\Dim-\Dik)\right]}\,\frac{c\tmax'}{3R'_i/R_i},\label{aprtm}\\
\fl \hbox{with:} \qquad \frac{c\tmax'}{3R'_i/R_i} = \left(\Dim-\frac{3}{2}\Dik\right)\,c(\tmax-\tbb)+\frac{c\tbb'}{3R'_i/R_i},\label{tmr}\ea
whereas as $L\to 0$ (as $t\to\tbb$ and $t\to\tcoll$) the leading terms are
\begin{equation}\HH_q\Dh \approx \mp\frac{\sqrt{m_{qi}}}{L^{3/2}} \to\mp\infty,\label{aprL0}\end{equation}
where $c(\tmax-\tbb)$ is given in terms of initial value functions by (\ref{tmc1})--(\ref{tmc2}) or (\ref{tmcOm}). The $\mp$ sign in (\ref{aprL0}) indicates that $\HH_q\Dh=\HH-\HH_q$ takes the opposite sign ($\pm$) of $\HH_q$ in the limits $L\to 0$, so that $0<\HH<\HH_q$ as $t\to\tbb$ (when $\HH,\,\HH_q\to\infty$) and $\HH_q<\HH<0$ as $t\to\tcoll$ (when $\HH,\,\HH_q\to -\infty$). This is consistent with $\Dh\to-1/2$ in both limits. 

Evidently, irrespectively of the choice of initial value functions and of the sign of $c\tmax'/R'_i$, the fact that $\HH_q\Dh$ evolves from $-\infty$ to $\infty$ implies that there must necessarily be a change of sign of $\HH'_q$, and so a profile inversion of $\HH$ is an inherent feature of open elliptic models that expand and collapse. We have at least the following two possibilities:
\begin{itemize}
\item If $\tmax'/R'_i\geq 0$, then $\HH_q\Dh>0$ as $t\to\tmax$, and so this quantity has necessarily changed sign at some $t<\tmax$. 

\item  If $\tmax'/R'_i\leq 0$, then $\HH_q\Dh<0$ has not changed sign as $t\to\tmax$. In this case, the change of sign of $\HH_q\Dh$ must occur at some $t>\tmax$, so that $\HH_q\Dh\to\infty$ as $t\to\tcoll$. 
\end{itemize}

\begin{figure}[htbp]
\begin{center}
\includegraphics[width=4.5in]{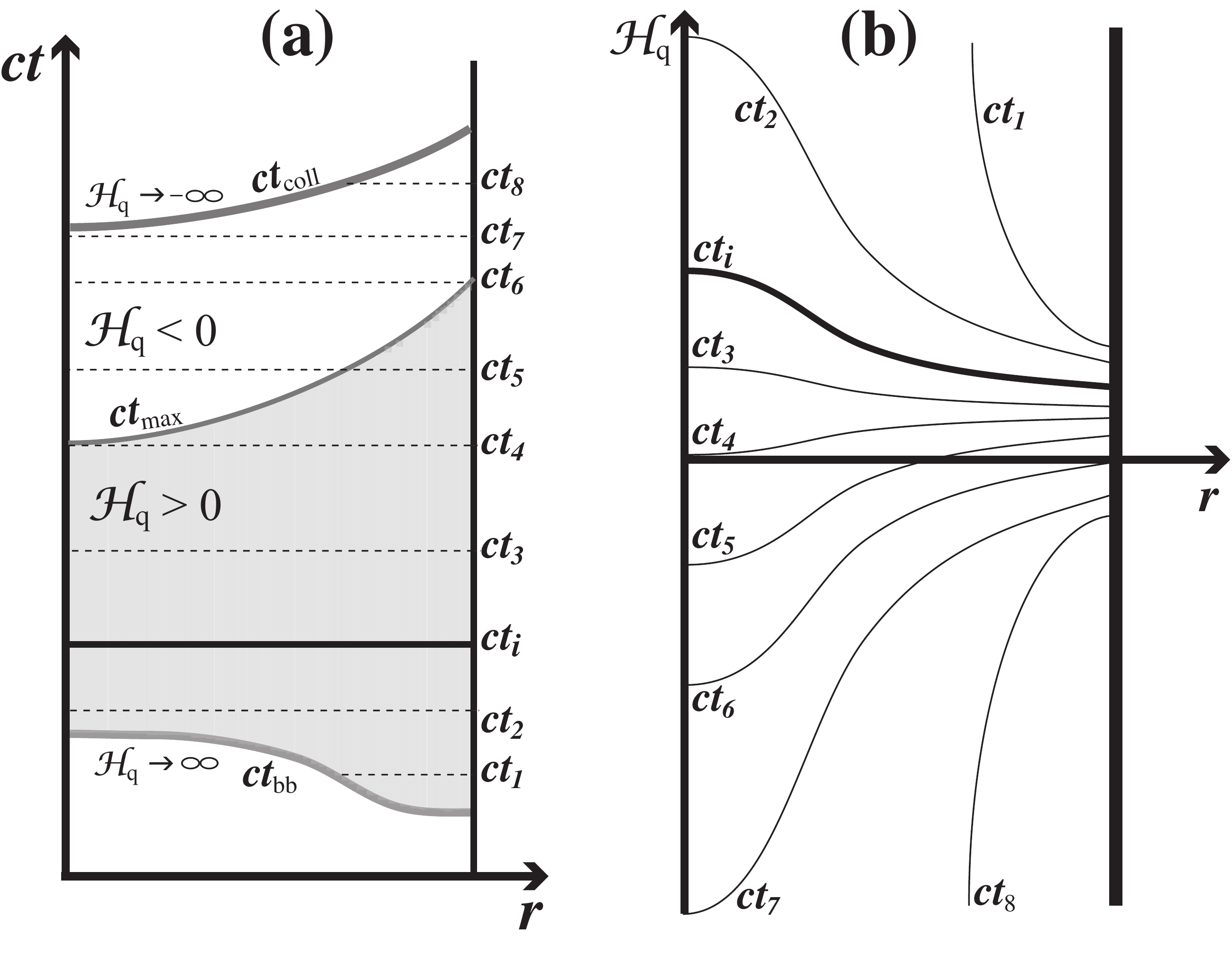}
\caption{{\bf Profile inversions of the expansion scalar in elliptic models.} Panel (a) displays the $(ct,r)$ plane associated with a radial domain (\ref{vartheta}). The region in which $\HH_q(ct,r)>0$ is shaded, so that its upper boundary marks the maximal expansion surface $t=\tmax(r)$, which coincides with $\HH_q(ct,r)=0$ and is not (in general) simultaneous. The locii of the initial big bang and final collapse singularities, $c\tbb(r)$ and $c\tcoll(r)$, are the lower and upper thick curves.  Panel (b) displays qualitative plots of the radial profiles of $\HH_q$ for the same values of $ct$ and $r$ shown in panel (a). Notice that the change of sign of $\HH_q$ and the fact that $\HH_q\to\pm\infty$ at the singularities strongly constrain these radial profiles. Since $c\tmax'\geq 0$, then (see section 8) the profile inversion has already taken place for $ct_4\leq c\tmax$ (where the equality only holds at $r=0$). From Lemma 3, the profiles of the local expansion $\HH$ qualitatively behave as those of $\HH_q$, but the surface $\HH=0$ does not coincide with $ct=c\tmax(r)$. In closed elliptic models $\HH_q$ and $\HH$ have a common TV with $R$ at $r=\rtv$ (not shown here). For these models the curves in both panels are qualitatively the same in the range $0\leq r< \rtv$, and are  mirror images reflected by a vertical axis for $\rtv < r< r_c$, where $r_c$ marks the second symmetry center (see section 9).}
\label{fig3}
\end{center}
\end{figure}

\noindent
In order to test these possibilities, we need to examine sign of $c\tmax'/R'_i$, which is easier to do by looking at the function $c\tmax(r)$ given by (\ref{tmc2}) and by considering the possible radial behavior of $x_i=k_{qi}/m_{qi}$ and $y_i=k_{qi}^{1/2}x_i$ in open elliptic models \cite{suss10b}. Considering (\ref{layer}) and using the coordinate gauge (\ref{rgauge}), $k_{qi}$ is restricted by
\bse\ba 1+E = 1- k_{qi}r^2\,R_0^2 > 0,\label{restrk1}\\
 0 < x_i <2,\label{restrk2}\ea\ese
which implies that (i) the limit $k_{qi}\to k_0 = \hbox{const.}>0$ as $r\to\infty$ is not possible, so that $k_{qi}$ must decay radially like $r^{-2}$ or faster, and (ii) if $m_{qi}\to 0$ then $m_{qi}$ must decay as fast $k_{qi}$, though $m_{qi}\to m_0 = \hbox{const.}>0$ is possible. This yields the following possibilities for the asymptotic limit of $x_i$ (see \cite{suss10b})
\bse\ba k_{qi}\to 0,\quad m_{qi}\to m_0,\qquad x_i\to 0,\quad y_i\to 0,\label{xias1}\\
   k_{qi}\to 0,\quad m_{qi}\to 0,\qquad x_i\to x_0\leq 2,\quad y_i\to 0.\label{xias2}
   \ea\ese
Notice that (\ref{xias1}) implies $x_i<2$, as $x\to 2$ is only possible if assuming (\ref{xias2}). If the asymptotic behavior is given by (\ref{xias1}) or (\ref{xias2}) with $x_0<2$, then $\pi-Z_e(x_i)$ remains positive and bounded, and thus $c\tmax(r)$ in (\ref{tmc2}) takes the radial asymptotic form
\begin{equation}c\tmax \approx ct_i+\frac{\pi}{y_i}\to\infty.\label{tmaxas1}\end{equation}
If we consider (\ref{xias2}) with $x_0=k_0/m_0=2$, then $x_i\approx 2+\epsilon$ were $\epsilon\sim O(r^{-\gamma})$, as we can assume that both $m_{qi}$ and $k_{qi}$ decay as $r^{-\gamma}$ with $2\leq \gamma\leq 3$ (see \cite{suss10b}). Expanding around $\epsilon=0$ we obtain
\begin{equation}c\tmax \approx ct_i+\frac{[\pi-Z_i(x_i)]r^{\gamma/2}}{2\sqrt{2m_0}}\approx ct_i+\frac{1}{\sqrt{m_0}}.\label{tmaxas2}\end{equation}
As a consequence of (\ref{tmaxas1}), if $m_{qi}$ and $k_{qi}$ behave asymptotically as in (\ref{xias1}) or (\ref{xias2}) with $x_0<2$, then we must necessarily have $c\tmax'>0$ at least in the asymptotic radial range, though $c\tmax'<0$ might still occur for values of $r$ closer to the center. However, in the case (\ref{xias2}) with $x_0=2$ it is impossible to assert the sign of $c\tmax'$ without making further assumptions on $m_{qi}$ and $k_{qi}$. Radial profiles and profile inversions of $\HH_q$ are displayed by figure \ref{fig3} for the case $c\tmax'\geq 0$. The profiles for the case  $c\tmax'= 0$ would be analogous to those of figure \ref{fig2} depicting a clump to void profile inversion without TV's, as in this case $\HH_q=0$ is simultaneous.  
\section{``Closed'' elliptic models.} 

So far we have only considered radial domains $\vartheta_c(r)$ defined by (\ref{vartheta}), containing a single symmetry center and complying with $R'>0$. The upper bound of these domains can be extended to encompass full ``open'' elliptic models whose hypersurfaces $\T(t)$ are homeomorphic to $\mathbb{R}^3$. If the hypersurfaces $\T(t)$ are homeomorphic to $\mathbb{S}^3$ ({\it i.e.} a ``closed'' model), then there is a second symmetry center $r=r_c$, so that the full radial range is $\{0\leq r\leq r_c\}$, and there is necessarily a TV of $R$ at $r=\rtv<r_c$ such that $R'(\rtv)=0$ occurs in regular conditions. 

All domains $\vartheta_c(r)$ with $r<\rtv$ in closed elliptic models are wholly equivalent to elliptic domains examined so far, hence the results proven in previous lemmas concerning elliptic models and regions remain valid. In particular, from Lemma 9 and the results of sections 6 and 7, the density and spatial curvature profiles in these domains must be monotonous and of a clump form for all $t$. However, domains $\vartheta_c(r)$ with $r>\rtv$ will contain a TV of $R$, and thus the results of the previous lemmas must be re--considered. We examine in the remaining of this section the properties of radial profiles in domains with $r>\rtv$. 

From the regularity condition (\ref{layer}), a well defined proper radial length (\ref{elldef}) requires the ratio $R'/\FF$ with $\FF=\sqrt{1+E}$ to be continuous (at least $C^0$) at $r=\rtv$. Since $R'(\rtv)=\FF(\rtv)=0$, with $R''(\rtv)<0,\,\FF'(\rtv)<0$, then from l'H\^oppital rule we have at the limit $r\to\rtv$
\begin{equation}  \frac{R'}{\FF}\to \frac{R''(\rtv)}{\FF'(\rtv)}>0, \label{RrFtv}\end{equation}
which implies that $\rtv$ is also a TV of the quasi--local volume element: $\VV_q'\dd r=4\pi R^2R'\dd r$, but not of the proper volume element: $\VV_p'\dd r=4\pi (R^2R'/\FF)\dd r$.  The spherical topology of the $\T[t]$ places a strong constraint on the radial profiles of scalars. We prove now the following

\begin{quote}

\noindent \underline{Lemma 10}. Let $A$ be any one of the covariant local scalars $\{m=\kappa\rho/3,\,k=\RR/6,\,\HH=\Theta/3\}$ associated with domain $\vartheta_c(r)$ with $r>\rtv$ in a closed elliptic model.  Then 
\begin{equation} R'(\rtv)=0\quad\Rightarrow\quad A'(\rtv)=0, \label{TVRT}\end{equation}
The converse statement of (\ref{TVRT}) is false: as we proved in previous lemmas, TV's of $m,\,k,\,\HH$ may exist under the assumption that $R'>0$ holds.\\   

\noindent \underline{Proof}.\,\,  By considering (\ref{propq2}) and (\ref{RrFtv}), it is evident that for all $A_q$
\begin{equation} R'(\rtv)=0\quad\Rightarrow\quad A_q'(\rtv)=0. \label{TVRTq}\end{equation}
To prove (\ref{TVRT}) we assume that $\rtv$ is a TV of $R$ and examine the behavior of $A$ and $A_q$ around $r=\rtv+\epsilon$ for $|\epsilon|\ll 1$. Considering (from (\ref{TVRTq})) that a TV of $R$ implies $A_q'(\rtv)=0$, we have at leading orders: 
\ba A_q(\rtv+\epsilon)\approx A_q(\rtv) +\frac{1}{2}A_q''(\rtv) \epsilon^2,
\nonumber\\
 R(\rtv+\epsilon)\approx R(\rtv) +\frac{1}{2}R''(\rtv) \epsilon^2, \nonumber\\\label{TRappr}\ea
Since, in general, $A(\rtv)\ne 0$, by applying (\ref{propq2}) to $A$ and $A_q$ and at leading orders we get
\ba \fl  A(\rtv+\epsilon)-A(\rtv)=A_q(\rtv+\epsilon)-A_q(\rtv)+\left[\frac{A_q' R}{3R'}\right]_{\rtv+\epsilon}-\left[\frac{A_q' R}{3R'}\right]_{\rtv}.\nonumber\\\ea
Using (\ref{TRappr}) to evaluate this expression at leading orders and dividing both sides by $\epsilon$ we get
\begin{equation} \fl \frac{A(\rtv+\epsilon)-A(\rtv)}{\epsilon}=\frac{A_q(\rtv+\epsilon)-A_q(\rtv)}{\epsilon}+\frac{1}{6}\,A_q''(\rtv)\,\epsilon.\end{equation}
Taking in both sides the limit as $\epsilon\to 0$ we get $A'(\rtv)=A_q'(\rtv)=0$, which is the desired result. \\

\noindent \underline{Corollary 1}.\,\, A TV of $R$ at $r=\rtv$ implies a TV of $M,\,E,\,\tbb,\,\tcoll$ and $\tmax$ at $r=\rtv$, but not a TV of $\FF$. Bearing in mind that a TV of $R$ in closed models holds for all $t$, it implies $R'_i(\rtv)=0$, the proof then follows directly from (\ref{tbbr}), (\ref{tcollr}), (\ref{scaling1a})--(\ref{scaling1b}) and (\ref{tmr}). On the other hand, a TV of $E$ does not imply a TV of $\FF$, which satisfies $1-\FF^2 = k_q R^2$. If we differentiate both sides of this expression and consider that the regularity condition (\ref{layer}) requires that $R'(\rtv)=0$ and $\FF(\rtv)=0$ hold, we find that $R'(\rtv)$ does not imply $\FF'(\rtv)=0$. \\

\noindent \underline{Corollary 2}.\,\, A TV of $R$ at $r=\rtv$ does not imply a change of sign of $\Da$ at $r=\rtv$. The proof follows directly from the definition of $\Da$ in (\ref{Dadef}) which involves the quotient of $A'_q$ and $R'$. At $r=\rtv$ the sign of $A'_q$ and $A'$ in the range $0<r<\rtv$ changes (Lemma 10) to its opposite sign in the range $\rtv<r<r_c$, but the sign of $R'$ changes at $\rtv$ in the same manner, thus keeping the sign of $\Da$ in $0<r<\rtv$ equal for the whole radial domain.       

\end{quote}

\noindent
As a direct consequence of Corollary 2, $\Dm\leq 0$ and $\Dk\leq 0$ must hold everywhere in all regular closed elliptic models. Since domains $\vartheta_c(r)$ with $r<\rtv$ are equivalent to domains considered by Lemma 9, then $\Dm\leq 0$ and $\Dk\leq 0$ must hold in these domains, but since the TV of $R$ at $r=\rtv$ does not change the signs of $\Dm$ and $\Dk$, then $\Dm\leq 0$ and $\Dk\leq 0$ must hold also for every domain with $r>\rtv$.

The common TV of scalars at the TV of $R$ (Lemma 10) marked by $\rtv$ is an inherent feature of the $\mathbb{S}^3$  topology of the $\T[t]$, and as such it is independent of the time evolution. Once we select an initial $\T[t_i]$ with this topology (which requires choosing $R_i$ such that $R'_i(\rtv)=0$), then $\rtv$ will be a common TV of the initial value functions $m_{qi},\,k_{qi}$ and (from Lemma 10) of $m_i,\,k_i$. The TV at $\rtv$ is then common to $m,\,m_q,\,k,\,k_q,\,\HH,\, \HH_q$. While there is no profile inversion of the density and spatial curvature because $\Dm\leq 0$ and $\Dk\leq 0$ must hold for all $t$, there is a local minimum of $m$ and $m_q$ at $\rtv$, and so we have a sort of ``topological density void'' in the form of an under--dense region marked by coordinate values around $r=\rtv$, between the two symmetry centers. This topological void is a sort of thick annular shell of under--dense dust layers concentric to both of the symmetry centers, and so it is of fundamentally different nature from the ``normal'' density void examined in previous sections, which occupies a closed compact under--dense region around a symmetry center. However, the most important difference is the fact that the normal density void is not a feature occurring at a fixed $r=\rtv$, but an evolved feature at changing values $r=\rtv[t]$ (the ``density wave'' \cite{kras1}) that depends on specific generic initial conditions at $t=t_i$, hence it always arises as the produce of a `clump to void' profile inversion.    

Since regular closed elliptic models exist that contain an arbitrary number of TV's of $R$ (as long as (\ref{layer}) holds at each TV), any of these configurations constitutes a complex pattern of topological clumps and voids in which the density profile changes from an over--density to an under--density and back at the TV's. However, as in the case of a single TV of $R$, these regions are concentric under--dense or over--dense thick spherical shells, not compact regions, and are fixed features (not evolving ``density waves'') that follow from the choice of an initial $\T[t_i]$ in which initial value functions $R_i,\,m_{qi},\,k_{qi}$ have an arbitrary number of common TV's. This inherent feature of closed elliptic models to produce TV's of density has been used in the literature  to generate void models by means of compound configurations constructed by suitably matching regions of closed elliptic models and sections of FLRW dust models \cite{cham,sato}. Since the voids and clumps in these configurations do not follow from generic initial conditions, they are wholly artificial and will not be considered any further.  

Regarding the profile of the expansion scalars $\HH$ and $\HH_q$, we take into consideration that the region $0\leq r< \rtv$ of any closed model is qualitatively analogous to a region of an open model containing a symmetry center. Therefore, the results of section 10 hold for such regions, which then necessarily have a profile inversion associated with a TV of $\HH_q$, which depending on the sign of $\tmax'$, will be either in the expanding or collapsing phase. From Lemma 3, there is necessary a profile inversion from a TV of $\HH$ as well. Since $\HH$ and $\HH_q$ must have a TV at $r=\rtv$, then radial profiles in the region $\rtv \leq r\leq  r_c$ are mirror images (reflected by $r=\rtv$) of the profiles in the region $0\leq r\leq \rtv$ (as those depicted by figure \ref{fig3}).   

\section{Special configurations.} 

\subsection{Models with a simultaneous big bang.}

Regular hyperbolic and elliptic LTB models exist for which the initial singularity occurs in a single singular $\T[t]$, hence $\tbb=t_0=\hbox{const.}<t_i$. We prove in this section that density and spatial curvature profile inversions are not possible for these configurations.  

Considering (\ref{xy}), (\ref{tbbhe}), (\ref{xye}), (\ref{tmc1}), (\ref{tbbr}), a simultaneous big bang implies the following constraints among the initial value functions
\bse\ba \fl c(t_i-t_0) = \frac{Z_h(x_i)}{y_i},\qquad \Dim =-\frac{\beta}{\alpha}\Dik,\quad \frac{\beta}{\alpha}>0,\qquad \hbox{hyperbolic},\label{simtbbh}\\
\fl c(t_i-t_0) = \frac{Z_e(x_i)}{y_i},\qquad \Dim =\frac{\beta_1}{\alpha_1}\Dik,\quad 0<\frac{\beta_1}{\alpha_1}<1,\qquad \hbox{elliptic},\label{simtbbe}\ea\ese
where $Z_h,\,Z_e$ are given by (\ref{hypZ1a}) and (\ref{ellZ1a}), while $\alpha,\,\beta,\,\alpha_1,\,\beta_1$ were defined by (\ref{tbbrh}) and (\ref{tbbre}). The constraints (\ref{simtbbh})--(\ref{simtbbe}) are clearly incompatible with (\ref{CintoV2}), hence there are no density profile inversions without a TV (turning value). 

The functions $\Gamma,\,\Dm$ and $\Dk$ in (\ref{Ghe}), (\ref{Dmposhe}) and (\ref{Dkposhe}) take the form
\ba\fl \Gamma = 1+3(\Dim-\Dik)\left(1-\frac{\HH_q}{\HH_{qi}}\right)-3\HH_q\,c(t-t_0)\,\left(\Dim-\frac{3}{2}\Dik\right),\label{simtbbG}\\
\fl \Dm = \frac{3}{\Gamma}\,\left(\Dim-\frac{3}{2}\Dik\right)\,\left[\HH_{qi}\,c(t_i-t_0)-\frac{2}{3}\right],\label{simtbbDm}\\
\fl \Dk = \frac{2}{\Gamma}\,\left(\Dim-\frac{3}{2}\Dik\right)\,\left[\HH_{qi}\,c(t_i-t_0)-1\right].\label{simtbbDk}\ea
The Hellaby--Lake conditions (\ref{noshxGh}) and (\ref{noshxGe}) reduce to \cite{suss10a}
\bse\ba \Dim> -1,\qquad \Dik>-\frac{2}{3}, \qquad \hbox{hyperbolic},\label{simtbbHLh}\\
\Dim > -1,\qquad \Dim-\frac{3}{2}\Dik> 0, \qquad \hbox{elliptic}.\label{simtbbHLe}\ea\ese
We look at the hyperbolic and elliptic cases separately below.
\begin{itemize}
\item {\underline{Hyperbolic models}}. We have $2/3<\HH_qc(t-t_0)<1$ with $\HH_qc(t-t_0)\to 2/3$ and $\HH_qc(t-t_0)\to 1$, respectively, as $L\to 0$ and $L\to\infty$. Hence, if the Hellaby--Lake conditions hold ($\Gamma>0$) the sign of $\Dm$ is the same as sign of $\Dim-(3/2)\Dik$, but from (\ref{simtbbh}) the sign of $\Dim$ must be opposite to the sign of $\Dik$, and also we have
\begin{equation}\Dim -\frac{3}{2}\Dik =-\left(\frac{\beta}{\alpha}+\frac{3}{2}\right)\Dik.\label{simtbbD32}\end{equation}
If $\Dim\leq 0$ and $\Dik\geq 0$, then $\Dim -(3/2)\Dik\leq 0$ follows from (\ref{simtbbD32}) and so $\Dm\leq 0$. If $\Dim\geq 0$ and $\Dik\leq 0$, then (\ref{simtbbD32}) yields $\Dm\geq 0$. We obtain analogous results for spatial curvature, since the sign of $\Dk$ is opposite to the sign of $\Dim -(3/2)\Dik\leq 0$, so that $\Dk$ has the same sign of $\Dik$. Therefore, there are no density or spatial curvature profile inversions, though a density void profile is possible for all $t>t_0$, as $\Dm\to 0$ as $L\to 0$ at the simultaneous big bang $t=t_0$.
\item {\underline{Elliptic models}}.  Since $\HH_qc(t-t_0)=0$ at $\tmax$, we have $0\leq \HH_qc(t-t_0)<2/3$ in the expanding phase and $\HH_qc(t-t_0)<0$ in the collapsing phase. The constraint (\ref{simtbbe}) implies that $\Dim$ must have the same sign as $\Dik$, hence the Hellaby--Lake conditions (\ref{simtbbHLe}) can only be fulfilled if $-1<\Dim\leq 0$ and $\Dik\leq 0$ hold. Following the same reasoning as in the hyperbolic case, it is easy to show that $\Dm\leq 0$ and $\Dk\leq 0$ must then hold for all $t$. Therefore, there are no density and spatial curvature profile inversions, and (as opposed to the hyperbolic case) density void profiles are not possible in regular configurations.    

\end{itemize}

\subsection{``Mixed'' models in which $k_q$ changes sign and models lacking a center.}

Regular LTB models can be constructed by glueing regions of hyperbolic, parabolic and elliptic types, either by matching the parts  at fixed comoving radii (as in \cite{boncham1,boncham2,meszaros,ltbstuff}), or simply by prescribing as part of the initial conditions a function $k_{qi}$ that changes sign along its radial domain. It is important to remark that matching conditions for this type of kinematically mixed regions allow for step discontinuities in the radial direction and on radial gradients of scalars. Thus, since  $\Dm,\,\Dk$ depend on the radial gradients $m'_q$ and $k'_q$, these relative fluctuations could be only piecewise continuous in these constructions (unless the sections are glued with the extra requirement that radial gradients are also continuous). 

The results proved in the lemmas in this article remain valid when applied separately to each region of  any compound configuration made by glueing any combination of parabolic, hyperbolic or elliptic regions complying with regularity conditions. This is obvious for any  region (for whatever sign of $k_q$) containing a center $r=0$, in which $\Da(t,0)=0$ must hold for all $t$. However, the lemmas are valid also for regions given by radial ranges $r_1 < r< r_2$ not containing a symmetry center, so that  $\Da$ need not vanish at the extremes $r_1,\,r_2$. In particular, we can even apply the results proven in the lemmas to LTB models that lack a symmetry center. In these cases (all of which are elliptic), the integral definition for quasi--local scalars (\ref{aveq_def}) may not converge, hence it is more convenient to define and compute $m_q,\,k_q,\,\HH_q$ first, and then use (\ref{qltransfm})--(\ref{qltransfH}) to obtain the local scalars $m,\,k,\,\HH$.  

As we show further below, a change of sign of $k_q$ at some $r=r_b$ introduces specific constraints on $\Dk$ that can affect the criterion for the existence of TV's and profile inversions. The possible combinations of glueing hyperbolic and elliptic regions with a parabolic region, either containing a center or not, force the constrain $k_q=0$ to hold in a given radial range, which is rather artificial, hence such combined constructions will not be examined here.    

\subsection{Mixed elliptic/hyperbolic construction.} 

Since an elliptic region enclosing a parabolic or hyperbolic region containing a center is incompatible with the Hellaby--Lake conditions \cite{bonnor85}, we consider the opposite construction made by an elliptic region containing a symmetry center surrounded by an expanding hyperbolic exterior. This interesting mixed configuration can be constructed by prescribing a given $m_{qi}>0$ together with 
\begin{equation}  k_{qi}\,\left\{ 
\begin{array}{l}
  >0 \quad \hbox{for}\quad 0\leq r<r_b,\quad\hbox{elliptic region}  \\
  =0 \quad \hbox{for}\quad r=r_b, \qquad\hbox{interface} \\
  <0 \quad \hbox{for}\quad r>r_b, \qquad\hbox{hyperbolic region}\\ 
 \end{array} \right..\label{ranges} \end{equation}  
where we assume that $k'_{qi}$ and $m'_{qi}$ are continuous (at least $C^1$) at $r=r_b$.  Since $k'_{qi}(r_b)<0$ and $k_{qi}(r_b)=0$, then
\begin{equation} \fl \quad \hbox{as}\quad r\to r_b\quad\hbox{we have}\quad \Dik\,\,\left\{ 
\begin{array}{l}
  \to-\infty \quad \hbox{for}\quad r<r_b,\quad\hbox{elliptic region}  \\
  \to\infty \quad \hbox{for}\quad r>r_b, \qquad\hbox{hyperbolic region}\\ 
 \end{array} \right.,\label{rangesDk} \end{equation}  
which implies that $\Dik>0$ must hold in the hyperbolic region at least in the neighborhood of $r= r_b$. From section 6 and Lemma 9, $\Dm\leq 0$ and $\Dk\leq 0$ must hold everywhere in the elliptic region. However, the diverging of $\Dik$ in (\ref{rangesDk}) does not allow us to test the Hellaby--Lake conditions (\ref{noshxGh}) and (\ref{noshxGe}) and to evaluate $\Dm$ and $\Dk$ at $r=r_b$ by setting $k_{qi}=0$ (or equivalently $x_i=0$) in (\ref{Ghe}), (\ref{Dmposhe}) and (\ref{Dkposhe}). Instead, since $r=r_b$ corresponds to $k_{qi}=0$, the form of these functions in the neighborhood of $r=r_b$ follows form power series expansions on $x_i=k_{qi}/m_{qi}=0$:
\bse\ba \HH_q\,c(t-t_i) &\approx& \frac{2}{3}\left(1-\frac{1}{L^{3/2}}\right)+O(x_i),\\
1-\frac{\HH_q}{\HH_{qi}} &\approx& 1-\frac{1}{L^{3/2}}+O(x_i), \ea\ese
so that $c\tbb',\,\Gamma,\,\Dm$ and $\Dk$ in  (\ref{Ghe}), (\ref{tbbr}), (\ref{Dmposhe}) and (\ref{Dkposhe}) take the forms
\bse\ba \frac{c\tbb'}{3R'_i/R_i} \approx \frac{2\Dim}{3\sqrt{m_{qi}}}+O(x_i),\label{tbbrmixed}\\
\Gamma \approx 1+\left(1-\frac{1}{L^{3/2}}\right)\Dim + O(x_i), \label{Gmixed}\\
\Gamma\Dm \approx \frac{\Dim}{L^{3/2}}+ O(x_i), \label{Dmmixed}\\
\Gamma\Dk \approx \Dik-\frac{2}{3}\Dim\left(1-\frac{1}{L^{3/2}}\right)+ O(x_i),\label{Dkmixed}\ea\ese
where all the terms $O(x_i)$ involve products of the form $k_{qi}\Dik$, which are finite even of $\Dik$ diverges. Therefore, $c\tbb',\,\Gamma$ and $\Dm$ change smoothly as $k_q$ passes from positive to negative in the interface $r=r_b$, and simply take their corresponding parabolic forms around the interface, so that regularity there is fulfilled if the parabolic Hellaby--Lake condition (\ref{noshx_par}) holds. Also, $\Dm<0$ necessarily holds in the interface and (by continuity) at least in its vicinity inside the hyperbolic region. From Lemma 9, the limits in (\ref{rangesDk}) and (\ref{Dkmixed}), it is evident that that $\Dk< 0$ holds for all times in the elliptic region, and in particular $\Dk\to -\infty$ holds in the limit $r\to r_b$ within  this region, while in the hyperbolic region we have $\Dk\to \infty$ as $r\to r_b$, and thus $\Dk> 0$ must hold at least in the neighborhood of $r=r_b$, all of which is consistent with $k'_q(t,r_b)<0$ and $k_q(t, r_b)=0$ holding for all $t$. 

As a consequence of the restrictions on $\Dk$ described above, there cannot be density or spatial curvature profile inversions in the elliptic region, in which $m_q$ and $k_q$ have  clump profiles (and thus $m$ and $k$ by Lemma 3). However, profile inversions of these scalars are possible in the hyperbolic region depending on the possibility to attain the right combinations of behavior of $\Dim$ and $\Dik$ that yield these inversions (as displayed in Table 1). 

For $r>r_b$ but close to $r_b$ we must have $\Dik>0$ and $\Dim< 0$, which corresponds to case (iii) in Table 1, so that (\ref{dm32dk}) does not hold. For these values of $r$ (close to $r_b$) there cannot be a density profile inversion, but a spatial curvature inversion must have happened for some $t<t_i$. Since $k_q<0$ holds for all $r>r_b$, then we must have $k_q\to-\infty$ as $t\to\tbb$ (or $L\to 0$) for every $r>r_b$, which means that $\Dk<0$ must have occurred for $r$ close to $r_b$ in hypersurfaces $\T[t]$ close to $\tbb$, but $\Dik>0$ for $t=t_i$, hence the profile inversion at $t<t_i$.\\ 

\noindent
For $r>r_b$ (but sufficiently far from $r_b$) in the hyperbolic region we can have the four combinations of profile evolutions outlined by Table 1:
\begin{itemize}
\item Case (iii). If $\Dik>0$ and $\Dim<0$ holds for all $r$ we have the same situation as in the layers close to $r_b$:\,\, $\Dm<0$ and $\Dk>0$ hold for $t>t_i$ and thus there are no profile inversions of $m_q$ and $k_q$ for $t>t_i$. In this case $\Dk>0$ for $k_q<0$ implies that $k'_q<0$ holds for for all $r$, with $k'_q\to 0$ possibly as $r\to\infty$, so that density decreases (clump) and spatial curvature becomes more negative as $r$ increases (void profile for negative $k_q$). 
\item Case (i). If $\Dik(r_1)=0$ with $r_1>r_b$ and becomes negative for all $r>r_1$, with $\Dim$ remaining negative for all $r$, then a density TV and density profile inversion will occur for some $t>t_i$ in the range $r>r_1$ if (\ref{dm32dk}) holds.
\item Case (ii). As above but (\ref{dm32dk}) does not hold in the range $r>r_1$, hence there is no density profile inversion for $t>t_i$, but there is a profile inversion of $k_q$.
\item Case (iv). As above, but $\Dim(r_2)=0$ with $r_2\geq r_1$ and becomes positive for all $r>r_2$. In this case $m_{qi}$ has already a void profile ($m'_{qi}\geq 0$) for $r\geq r_2$, and since (\ref{dm32dk}) holds this profile is kept for the whole evolution (though a density  profile inversion must have occurred for some $t<t_i$).     
\end{itemize}
Evidently, the initial value functions $m_{qi},\,k_{qi},\,\Dim$ and $\Dik$ must also comply with the regularity conditions (\ref{noshxGh}), which do not follow from the qualitative restrictions described above.     

\section{Summary and conclusion.}

We have conducted a comprehensive and rigorous examination of the radial profiles of the main covariant scalars (density, spatial curvature and expansion, respectively denoted by $m,\,k,\,\HH$, see definition (\ref{mkHdefs})) for regular LTB models in full generality. The time evolution of these profiles has been considered, and in particular, we have addressed the issue of profile inversions in which an initial clump can evolve into a void (or vice versa). The necessary background material is the formalism of quasi--local variables within an initial value formulation (section 3), rephrasing in terms of this formalism the analytic solutions (Appendix B) and the Hellaby--Lake conditions (Appendix C), and the relation between the radial coordinate and proper radial length (Appendix D). The quasi--local density ($m_q$), spatial curvature ($k_q$) and expansion scalar ($\HH_q$) are given by \ref{mq})--(\ref{Hq}), while the fluctuations $\Da$ follow from (\ref{Dadef}).  

\subsection{Summary of results.}

We have used thoughout the article the term ``turning value'' of any scalar $A$ (``TV of $A$'') to denote a value $r=\rtv$ such that $A'$ vanishes at $r=\rtv$ at a given hypersurface $\T[t]$ marked by constant $t$ (this definition also applies to the $A_q$). In general, the value $\rtv$ is different for each $\T[t]$, but it can be the same for all $t$ (as for example a regular TV of $R$). Throughout sections 3--8 we have considered only LTB models or regions containing a symmetry center and with an open topology ($R'>0$ holds everywhere: no TV of $R$), leaving the case with closed topology (regular TV of $R$) and special configurations for sections 9 and 10. The main results contained in 10 lemmas that were proven in these sections are summarized below: 

\begin{itemize}   

\item {\bf{Section 3}}. The clump/void character of radial profiles of local and quasi--local scalars, $A$ and $A_q$, in radial domains containing a center was given in terms of the existence of turning values, TV's. Assuming that $R'>0$ holds, we proved the following lemmas: 
\begin{itemize}
\item {\bf{Lemma 1}}: The signs of the radial gradient $A'_q$ is related to the sign of the fluctuation $\Da$.
\item {\bf{Lemma 2}}: The necessary and sufficient condition for a TV of $A_q$ at some $r=\rtv$ is given by $\Da=0$ at $r=\rtv$. In general, $\rtv=\rtv[t]$ is different at different times (see figure 1). 
\item {\bf{Lemma 3}}: The existence of a TV of $A_q$ in a radial domain is a sufficient condition for the existence of a TV of $A$ in the same domain. 
\end{itemize}
These lemmas allow us to examine the monotonicity of scalars by looking at signs and the zeros of their fluctuations. As a consequence of Lemma 3, it is sufficient to examine the profiles of the scalars $A_q$ (which satisfy simpler scaling laws) to know the profiles of the scalars $A$.  

\item {\bf{Section 4}}. Character of initial density and spatial curvature profiles compatible with the Hellaby--Lake conditions. Initial density must have a clump profile in parabolic and elliptic models or regions, while initial clump and void profiles are possible in hyperbolic models or regions. Initial curvature  must have a clump profile in elliptic models or regions, but hyperbolic models or regions admit initial clumps and voids.  Explicit conditions are given for each case.     

\item {\bf{Section 5}}. Formal definition of profile inversion and its relation with TV's of scalars. Profile inversions can occur with and without TV's. The following lemmas were proven:
\begin{itemize}
\item {\bf{Lemma 4}}: A profile inversion of $A_q$ implies a profile inversion of $A$.
\item {\bf{Lemma 5}}: Sufficient conditions for the existence of a profile inversion of $A_q$ follow from the existence of TVs of $A_q$ plus some extra requirements on the fluctuations $\Da$. This type of profile inversion is illustrated by figure \ref{fig1}.
\item {\bf{Lemma 6}}: Sufficient conditions for the existence of a profile inversion of $A_q$ when there are no TV's of $A_q$. This type of profile inversion is illustrated by figure \ref{fig2}.
\end{itemize}
Notice that Lemma 4 guarantees that the results of Lemmas 5 and 6 apply also to the local scalars $A$. The conditions in Lemmas 5 and 6 are independent of the Hellaby--Lake conditions, hence the latter place extra constraints when applying these lemmas to specific configurations. 
\item {\bf{Section 6}}. Density radial profiles and profile inversions are examined in detail. The following lemmas were proven:
\begin{itemize}
\item {\bf{Lemma 7}}: Necessary and sufficient condition are given for the `clump to void' density profile inversion with density TV's in regular hyperbolic models or regions. Void to clump inversion is not possible. 
\item {\bf{Lemma 8}}: Necessary and sufficient condition for the `clump to void' inversion without TV's in regular hyperbolic models or regions. Void to clump inversion is not possible.
\item {\bf{Lemma 9}}: Density profile inversions, with or without density TV's, cannot occur in regular open elliptic models or regions containing a center (no TV's of $R$). 
\end{itemize}
The main result of this section (and possible of the whole article) follows as a consequence of these lemmas: 

\begin{quote}
\noindent
{\it {the only density profile inversion (with and without TV's) compatible with absence of shell--crossings (the Hellaby--Lake conditions) is the `clump to void' inversion in hyperbolic models or regions.}}
\end{quote}

\noindent 
The specific conditions outlined by Lemmas 7 and 8 (equations (\ref{dm32dk}) and (\ref{CintoV2})), which must be satisfied by hyperbolic models and region, are simple restrictions on their initial value functions (density, spatial curvature and their fluctuations). It is evident that these are not weird or outlandish conditions, but reasonable and easy to prescribe.  
\item {\bf{Section 7}}. Spatial curvature profiles and their inversions. The only inversion compatible with the Hellaby--Lake conditions is the `clump to void' inversion in hyperbolic models or regions. These inversions only occur in the particular parameter cases when there is no inversion of the density profile. 
\item {\bf{Table 1}} provides the combination of initial conditions that allow for profile inversions of density and spatial curvature in hyperbolic models or regions.
\item {\bf{Section 8}}. Radial profiles and profile inversions of the expansion scalar. Hyperbolic models and regions allow for a regular evolution with and without profile inversions (conditions are provided for each case). However, profile inversions of the expansion scalar necessarily occur in elliptic models or regions containing a center. This is illustrated by figure \ref{fig3}. 

\item {\bf{Section 9}}. Closed elliptic models. A TV of $R$ occurs under regular conditions at the same value $r=\rtv$ for all times. We proved the following lemma:
\begin{itemize}
\item {\bf{Lemma 10}}: The TV of $R$ at $r=\rtv$ implies that $\rtv$ is a common TV of the local and quasi--local density, spatial curvature and expansion scalar. The converse is not true, as we proved in Lemmas 1--5 and 7--8 that TV's of these scalars can occur when there is no TV of $R$ ($R'>0$ holds everywere). 
\end{itemize}
As an important corollary of this lemma, the fluctuations $\Da$ do not change sign because of the TV of $R$. As a consequence, there are no profile inversions of density and spatial curvature, though profile inversions of the expansion scalar must occur as in open elliptic models.   
\item {\bf{Section 10}}. Special configurations:
\begin{itemize}
\item {\underline{Regular hyperbolic and elliptic models with a simultaneous big bang}}. Profile inversions of density and spatial curvature (with or without TV's) cannot occur. Hyperbolic models allow for a void profile for all the time evolution. 
\item {\underline{Regions not containing symmetry centers}}. The results of all proven lemmas hold, the only caveat being the fact that (in general) we have $\Da\ne 0$ at the boundaries of the radial range (as opposed to $\Da= 0$ strictly holding at the symmetry center).
\item {\underline{LTB models with ``mixed'' kinematics}}. These configurations are made by glueing or matching combinations of parabolic, hyperbolic and elliptic regions (see \cite{boncham1,boncham2,meszaros,ltbstuff}). The results of all proven lemmas hold. 
\item {\underline{Elliptic region surrounded by a hyperbolic exterior}}. From Lemma 9 and assuming that Hellaby--Lake conditions hold, the elliptic region must have clump profiles for the density and spatial curvature, without inversions. However, void profiles and `clump to void' inversions are possible in the hyperbolic exterior in agreement with Lemma 7 and the combination of initial conditions specified in Table 1. 
\end{itemize}

\end{itemize}

\subsection{Relaxation of regularity conditions.}

In obtaining the results summarized above we have assumed that the Hellaby--Lake conditions \cite{HLconds,ltbstuff,suss10a} (see Appendix C) hold for all the evolution times of the models (shell crossings are completely absent). Evidently, if we relax these conditions by demanding that they hold only for all $t>t_i$ for a given $t=t_i$, then some of the results of the Lemmas are no longer binding. The most important example is Lemma 9, which forbids density void profiles and `clump to void' density profile inversion in elliptic models. This result follows from the fact that a density void profile is incompatible with the joint fulfillment of two of the Hellaby--Lake conditions (\ref{noshxGe}):\,  $\tbb'\leq 0$ and $\tcoll'\geq 0$, where $\tbb$ and $\tcoll$ denote the locus of the initial and collapsing singularity. However, if we demand $\tcoll'\geq 0$ to hold but not $\tbb'\leq 0$, then a shell crossing singularity necessarily emerges for $t\approx \tbb$ for all $r$, but not for $t\approx \tcoll$. Depending on the free parameters $m_{qi},\,k_{qi}$, it may be possible to construct an elliptic model that is free from shell crossings for all $t>t_i$ for some $t_i>\tbb$ \cite{suss02}, and that admits a `clump to void' inversion and density void profiles in this time range. Since a dust source does not provide a good description of the physical conditions near an initial singularity in a cosmological model, it may be sufficient for a physically reasonable evolution to demand absence of shell crossings for all $t>t_i$, provided one can justify that the dust description beaks down for $t<t_i$. 

The relaxation of the Hellaby--Lake conditions, as described above, can also work for hyperbolic models or for special configurations. It can also allow for a density `void to clump' profile inversion, and it can also be set up so that an evolution free from shell crossings occurs in time ranges restricted to an early stage $\tbb < t<t_i$. This latter case can be useful for studying void formation in the context of toy models of structure formation (``spherical collapse model''), in which the collapsing stage of an over--density with positive spatial curvature (elliptic) is omitted,  as the late stage of the dynamics is dominated by the virialization process leading to the formation of stable structures (hence  there is no collapse stage because the assumption of spherical dust is no longer valid)\cite{padma}.          

\subsection{Final discussion.}

The most important results in this article are those obtained in section 6 (Lemmas 7, 8 and 9), section 9 (Lemma 10 and its corollaries) and the elliptic/hyperbolic mixed configuration in section 10. These results correct and provide full generality to the work initiated by Mustapha and Hellaby \cite{mushel}, as these authors only furnished a restricted proof (together with particular and numeric examples) of the existence of the `clump to void' density profile inversion in hyperbolic models, without attempting to prove it for elliptic or parabolic models. They also claimed (mistakenly) that the `void to clump' inversion was possible without shell crossings. Since they only examined density profiles, the results of sections 7 and 8 also extend and generalize their work to spatial curvature and the expansion scalar. 

The fact that the existence of density void profiles (with total absence of shell crossings) is only possible in hyperbolic models and regions is consistent with the conclusions of \cite{boncham1,boncham2,meszaros}. However, as opposed to these authors, we have obtained the analytic conditions for these profiles and also for their inversions in full generality, and without resorting to artificial configurations made by matchings regions in which the free parameters take special forms and radial gradients and density are likely discontinuous.  Since the results of Lemma 9 are analytic and general, the elliptic models with void profiles examined in \cite{occhio81} must necessarily have shell crossings (which was not proven otherwise by the authors). Also, some of the configurations in the first and second column of Table 1 of \cite{meszaros} are listed as free from shell crossings. This is mistaken, as the Hellaby--Lake conditions are incompatible with elliptic and parabolic regions that contain a symmetry center and have density void profiles.   

Radial profiles of scalars in closed elliptic models are strongly constrained by the TV of $R$ that occurs at a fixed value $r=\rtv$, which (as proven by Lemma 10) is a common TV for the density, spatial curvature, expansion scalar and all initial value functions (but not the initial fluctuations). Since this TV implies a change of sign of the gradients $m'_q$ and $m'$ at $r=\rtv$, it introduces relative under--dense or over--dense regions in thick spherical shells of dust layers, as opposed to a compact spherical region around the center. These shells can be considered as some sort of  ``topological'' density clumps or voids (the same remark applies to spatial curvature or the expansion scalar), and are a product of the spherical $\mathbb{S}^3$ topology of the rest frames $\T[t]$, and as such are not related to a profile inversion produced by the evolution of the models given generic initial conditions (as is the case for the TV's that were examined in sections 3--8). As a consequence, models of topological clumps/void profiles \cite{cham,sato} are rather artificial constructions.

It is important to emphasize that the radial profile of the expansion scalar, as examined in section 8, is much less restrictive to the existence of TV's and profile inversions than the profiles of density and spatial curvature. This fact is crucial to assess the existence and magnitude of a ``back--reaction'' term and an ``effective'' acceleration in the context of Buchert's scalar averaging applied to LTB models \cite{LTBave2,LTBave3}, which is a relevant issue in current research involving LTB models. Also, as pointed out in recent literature \cite{wiltshire}, spatial gradients of the expansion scalar (related to gradients of the binding energy) are the main generators of back--reaction (see also \cite{LTBave2,LTBave3}), and as such are important in providing a theoretical interpretation of cosmological observations that avoid introducing a dark energy source.        

Regarding the elliptic region surrounded by a hyperbolic external region, the fact that the latter may admit a density void profile (and a `clump to void' inversion), without violating regularity conditions (Hellaby--Lake), is indeed a nice and unexpected result. This possibility leads to simple models in which a high density region undergoing collapse is surrounded by a large spherical shell in an intermediary transition scale, whose density is lower than that of the cosmic large scale asymptotic background. Configurations of this type could be compatible with the notion of ``finite infinity'' in which cosmic bound structures are approximately regarded as asymptotically flat at such intermediate scales between the cosmic voids and the far cosmic background \cite{wiltshire,fiEllis}. 

Finally, it is necessary to remark that we have conducted a theoretical study of radial profiles and profile inversions based on general analytic expressiond (avoiding the use of special cases based on narrow parameter specializations). Hence, we have not taken into consideration the question of whether these radial profiles are compatible or not with the constraints that follow from actual cosmological and astrophysical observations. However, introducing specific profiles meeting these observational criteria is an important task that certainly requires a follow up article based on a full numeric approach, preferably using the Omega and Hubble parameters discussed in Appendix E. This article is presently under elaboration.

\begin{appendix}
\section{Analytic solutions and the Hellaby--Lake conditions in the conventional variables. }

The solutions of the Friedman--like field equation (\ref{fieldeq1}) for each kinematic class take the following well known parametric form:\\

\noindent
{\underline{Parabolic models or regions:}} \,\, $E=0$.
\begin{equation}c(t-\tbb) = \frac{2}{3}\,\eta^3,\qquad R=(2M)^{1/3}\,\eta^2,\label{par1}\end{equation}
\noindent
{\underline{Hyperbolic models or regions:}} \,\, $E\geq 0$.
\begin{equation} 
R =\frac{M}{E}\,\left(\cosh\,\eta-1\right),\qquad 
c(t-\tbb)=\frac{M}{E^{3/2}}\,\left(\sinh\,\eta-\eta\right),\label{hypt1}
\end{equation}
\noindent
{\underline{Elliptic models or regions:}} \,\, $E\leq 0$.
\begin{equation}
R =\frac{M}{|E|}\,\left(1-\cos\,\eta\right),\qquad
c(t-\tbb)=\frac{M}{|E|^{3/2}}\,\left(\eta-\sin\,\eta\right),\label{ellt1}          
\end{equation}
where $\tbb=\tbb(r)$ is called ``big bang time'', as it marks the coordinate locus of the central expanding curvature singularity: $R(t,r)=0$ for $r\geq 0$. This free function emerges as an ``integration constant''  in the integration of (\ref{fieldeq1}). Notice that the locus of central curvature singularity is distinct from that of the center of symmetry $R(t,0)=0$ (see Appendix A1 of \cite{suss10a}). Besides the kinematic class, LTB models that admit (at least) one symmetry center can be classified as ``open'' or ``closed'', respectively corresponding to the hypersurfaces of constant $t$ being topologically equivalent to $\mathbb{R}^3$ or $\mathbb{S}^3$ (see Appendix D and Appendix A3 of \cite{suss10a}).

\section{The analytic solutions in terms of an initial value approach. }

The Friedman--like equation (\ref{fieldeq1}) is equivalent to (\ref{Hq}), hence the analytic solutions of the former (given by (\ref{par1})--(\ref{ellt1})) are equivalent to those of the latter, but given in terms of $L$ and the initial value functions $\{m_{qi},\,k_{qi}\}$, which are related to the free parameters $M$ and $E$ by (\ref{ME}). Notice that $\{m_{qi},\,k_{qi}\}$ is an irreducible set of free functions, since $R_i$ can be specified as a radial coordinate gauge. As we show below, the ``bang time'', $\tbb(r)$,  follows as a function of $m_{qi},\,k_{qi}$. See Appendix E for an alternative set of initial value functions that can be related to the Hubble and Omega factors of a FLRW cosmology.  

\subsection{Parabolic models or regions: $k_{qi}=0$}

We express $M$ and $R$ in terms of $m_{qi}$ and $L$ by inserting (\ref{Ldef}) and (\ref{ME}) in (\ref{par1}). After re--arranging terms we get the following closed analytic expression for $L$
\begin{equation} L =\left[1+\frac{3}{2}\sqrt{2m_{qi}}\,c(t-t_i)\right]^{2/3},\label{par2}\end{equation}
where we are only considering expanding configurations ($L$ increases for $t>t_i$). The bang time follows by considering that $L_i=1$ and setting $L=0$ and $t=\tbb$ in (\ref{par2})
\begin{equation}  c\tbb = ct_i-\frac{2}{3\sqrt{2m_{qi}}}=ct_i-\frac{2}{3\HH_{qi}}.\label{tbbpar}  \end{equation}

\subsection{Hyperbolic models or regions: $k_{qi}< 0$}

We obtain the following implicit solution of the form $t=t(R)$ by eliminating the parameter $\eta$ from the equation for $R$ in (\ref{hypt1}) and substituting in the equation for $t$:
\begin{equation} \frac{E^{3/2}}{M}\,c(t-\tbb)=Z_h(\bar R),\label{hypZR}\end{equation}
where $\bar R= (E/M)R$ and $Z_h$ is the function 
\begin{equation} u  \mapsto Z_h(u)=u ^{1/2} \left( {2 + u } \right)^{1/2}  - \hbox{arccosh}(1 + u ).\label{hypZ1a}  \end{equation}
We express then $M,\,E$ and $R$ in (\ref{hypZR}) in terms of  $m_{qi},\,k_{qi}$ and $L$ from (\ref{Ldef}) and (\ref{ME}). The result is
\begin{equation} y_i\, c(t-t_i) = Z_h(x_iL) - Z_h(x_i),\label{hypZ2}\end{equation}
where
\begin{equation}
 x_i = \frac{|k_{qi}|}{m_{qi}},\qquad y_i=\frac{|k_{qi}|^{3/2}}{m_{qi}}.\label{xy}
\end{equation}
Setting $L_i=1$, \,$t=\tbb$ and $L=0$ in (\ref{hypZ2}) and using (\ref{hypZ1a}) yields the bang time as a function of $m_{qi}$ and $|k_{qi}|$:
\begin{equation} c\tbb = ct_i-\frac{Z_h(x_i)}{y_i}.\label{tbbhe} \end{equation}

\subsection{Elliptic models or regions: $k_{qi}> 0$}

In this case, the implicit solution $t=t(R)$ follows from eliminating $\eta$ from the first equation in (\ref{ellt1}) and substituting in the second one. The resulting implicit solution  has two branches, an ``expanding'' one ($0<\eta<\pi$ with $\dot R>0$) and a ``collapsing'' one ($\pi<\eta<2\pi$ with $\dot R<0$):
\begin{equation} \frac{|E|^{3/2}}{M}\,c(t-\tbb)=\left\{ \begin{array}{l}
  Z_e(\bar R)\qquad\quad\; \hbox{expanding phase} \\ 
  2\pi-Z_e(\bar R) \quad \hbox{collapsing phase} \\ 
 \end{array} \right.,\label{ellZR} \end{equation}
where $\bar R= (|E|/M)R$ and $Z_e$ is given by
\begin{equation} u  \mapsto  Z_e(u)= \arccos(1 - u )-u ^{1/2} \left( {2 - u } \right)^{1/2}.\label{ellZ1a}  \end{equation}
Proceeding as in the hyperbolic case, we transform (\ref{ellZR}) into
\begin{equation}  y_i\, c(t-t_i)+Z_e(x_i)  = \left\{ \begin{array}{l}
 Z_e(x_iL) \qquad\qquad \hbox{expanding phase}\\ 
  \\ 
 2\pi-Z_e(x_iL) \qquad \hbox{collapsing phase}\\ 
 \end{array} \right.\label{ellZ2}\end{equation}
where
\begin{equation}
 x_i = \frac{k_{qi}}{m_{qi}},\qquad y_i=\frac{k_{qi}^{3/2}}{m_{qi}},\label{xye}
\end{equation}
It follows from (\ref{ellZ2})--(\ref{ellZ1a}) that $L$ is restricted by $0<L\leq \Lmax$, where the maximal expansion is  $\Lmax=2/x_i=2m_{qi}/k_{qi}$, characterized by  $\dot L=0$ and $\HH_q=0$. 

Setting $L_i=1$,\, $t=\tbb$ and $L=0$ in the expanding phase of (\ref{ellZ2}) yields a bang time function, while $t=\tcoll$ and $L=0$ in the collapsing phase yields the ``crunch'' time associated with the collapsing singularity. The maximal expansion time follows by substituting  $t=\tmax$ and $L=\Lmax$ in either branch of (\ref{ellZ2}). These times are given by
\ba  c\tbb &=& ct_i-\frac{Z_e(x_i)}{y_i},\label{tmc1}\\
 c\tmax &=& c\tbb+\frac{\pi}{y_i}=ct_i+\frac{\pi-Z_e(x_i)}{y_i},\label{tmc2}\\  c\tcoll &=& c\tbb+\frac{2\pi}{y_i}=ct_i+\frac{2\pi-Z_e(x_i)}{y_i}\label{tmc3}.\ea
Notice that (in general) $\tmax=\tmax(r)$ and $\tcoll=\tcoll(r)$, like $\tbb(r)$, are not simultaneous (do not coincide with a $\T[t]$ hypersurface). For every comoving observer $r =$ const., the time evolution is contained in the range $\tbb(r)<t<\tcoll(r)$.

The parametrization of initial conditions given here in terms of $\{m_{qi},\,k_{qi}\}$ is not unique. We could have used $\{m_{qi},\,\HH_{qi}\}$ as initial value functions, so that  $k_{qi}=2m_{qi}-\HH_{qi}^2$ follows from (\ref{Hq}). An alternative description (which can be suitable for applications) is provided in Appendix E in terms of initial value functions that resemble observational parameters (the Hubble and Omega parameters of a FLRW cosmology).

\section{Avoidance of shell crossings: the Hellaby--Lake conditions.}

The restrictions on the free parameters $\{M,\,E,\,c\tbb\}$ that guarantee an evolution free from shell crossings (thus complying with (\ref{noshxG})) are the well known Hellaby--Lake (necessary and sufficient) conditions \cite{HLconds,ltbstuff,suss02} given for each kinematic class by\\ 

\noindent {\underline{Parabolic and hyperbolic models or regions:}}
\begin{equation} R'>0\quad \Leftrightarrow\quad \left\{ M'\geq 0,\quad E'\geq 0,\quad \tbb'\leq 0\right\},\label{noshx1}\end{equation}

\noindent{\underline{Elliptic models or regions:}}
\ba  \pm R' > 0\quad &\Leftrightarrow& \quad \pm  M'\geq 0,\quad \pm \tbb'\leq 0,\nonumber\\
&{}& \quad\pm\left[\frac{M'}{M}-\frac{3}{2}\frac{E'}{E}+\frac{c\tbb'\,|E|^{3/2}}{2\pi\,M}\right]\geq 0,\nonumber\\\label{noshx2}
\ea
where only expanding configurations are considered in (\ref{noshx1}) and the $\pm$ sign in (\ref{noshx2}) accounts for the fact that $R'<0$ occurs in elliptic models whose that admit a second symmetry center (see section 9 and Appendix A3 of \cite{suss10a}). The equal sign holds only at symmetry centers and at values of $r$ where $R'=0$.

In order to express the Hellaby--Lake conditions in terms of the quasi--local scalars in the initial value formulation  
\footnote{We examine the Hellaby--Lake conditions in the parametrization $\{m_{qi},\,k_{qi}\}$, as it is straightforward to re--phrase all expressions in terms of the initial conditions $\{\HH_{qi},\,\hOmi\}$ defined in Appendix E}
we examine the sign condition of the analytic forms of $\Gamma$ in (\ref{Gp}) and (\ref{Ghe}) \cite{suss10a}. A comparison with the forms (\ref{noshx1}) and (\ref{noshx2}) follows by expressing the gradients $M',\,E'$ and $c\tbb'$ in terms of our initial value functions by means of
\begin{equation} \frac{M'}{M}=\frac{3R'_i}{R_i}\left[1+\Dim\right],\qquad \frac{E'}{E}=\frac{3R'_i}{R_i}\left[\frac{2}{3}+\Dik\right],\label{MEDi}\end{equation}
\ba  \frac{c\tbb'}{3R'_i/R_i} &=& c(t_i-\tbb)\,\Dim =\frac{2\,\Dim}{3\sqrt{2m_{qi}}},\qquad \hbox{parabolic}\label{tbbrp}\\ 
 \frac{c\tbb'}{3R'_i/R_i} 
&=& \frac{\Dim-\Dik}{\HH_{qi}}-c(t_i-\tbb)\left(\Dim-\frac{3}{2}\Dik\right),\nonumber\\&{}& \hbox{hyperbolic and elliptic}\label{tbbr}\ea
where we differentiated both sides of (\ref{ME}), (\ref{tbbpar}) and (\ref{tbbhe}) with respect to $r$ and used (\ref{Dadef}) specialized to $t=t_i$, while $\tbb$ in (\ref{tbbr}) is given by (\ref{tbbhe}) with $Z_h(x_i)$ or $Z_e(x_i)$, respectively, for hyperbolic and elliptic models. Considering (\ref{MEDi}) and (\ref{tbbrp})--(\ref{tbbr}), the Hellaby--Lake conditions for each kinematic class are: 

\begin{itemize}
\item {\underline{Parabolic models or regions.}}
\begin{equation} -1\leq \Dim\leq 0,\label{noshx_par}\end{equation}      
which from (\ref{Dadef}) implies $m'_{qi}\leq 0$. It is evident from (\ref{MEDi}) and (\ref{tbbrp}) that (\ref{noshx_par}) implies $\tbb'/R'_i\leq 0$ and $M'\geq 0$, and so it is equivalent  to the Hellaby--Lake conditions (\ref{noshx1}).

\item {\underline{Hyperbolic models or regions.}}
\begin{equation} c\tbb'\leq 0,\qquad \Dik\geq -\frac{2}{3},\qquad \Dim\geq -1.\label{noshxGh}\end{equation}
where $\tbb'$ is given in terms of our initial value functions by (\ref{tbbr}). Notice, from (\ref{slawDk}), that all regular hyperbolic models comply with $\Dk\geq -2/3$  for all their time evolution. It is evident from (\ref{MEDi}) that (\ref{noshxGh}) are completely equivalent to the Hellaby--Lake conditions (\ref{noshx1}).

\item {\underline{Elliptic models or regions.}}
\begin{equation} \frac{c\tbb'}{3R'_i/R_i}\leq 0,\qquad  \frac{c\tcoll'}{3R'_i/R_i}\geq 0,\qquad \Dim\geq -1,\label{noshxGe}\end{equation}
with
\begin{equation} \frac{c\tcoll'}{3R'_i/R_i} =\left(\Dim-\frac{3}{2}\Dik\right)\,c(\tcoll-\tbb)+\frac{c\tbb'}{3R'_i/R_i},\label{tcollr}\end{equation}
where we used $c(\tcoll-t_i)=c(\tcoll-\tbb)-c(t_i-\tbb)$, with $\tbb$ and $\tcoll$ given by (\ref{tmc1}) and (\ref{tmc3}), and the equal sign holds only at a symmetry center. Notice that two of the three conditions in (\ref{noshxGe}) are sign conditions on the gradients of the coordinate locus of the central singularity ($\tbb',\,\tcoll'$). This fact has not been, apparently, noticed in previous literature \cite{kras1,kras2,ltbstuff,suss02}. 

Conditions (\ref{noshxGe}) imply the following necessary (but not sufficient) condition for (\ref{noshxG})
\begin{equation}\Dim-\frac{3}{2}\Dik\geq 0,\label{noshx_nec1}\end{equation}
which, from (\ref{slawD32}), implies that $\Dm-(3/2)\Dk\geq 0$ necessarily holds for all times if (\ref{noshxG}) holds. Since condition (\ref{noshx_nec1}) implies
\begin{equation} 1+3(\Dim-\Dik)\ge 0,\label{Gmax},\end{equation}
this (also necessary but not sufficient) condition follows from (\ref{noshxGe}) too. It is straightforward to show from (\ref{Dadef}), (\ref{tmc1}), (\ref{tmc3}), (\ref{MEDi}), (\ref{tbbr}) and (\ref{tcollr}) that conditions (\ref{noshxGe}) are equivalent to the Hellaby--Lake conditions (\ref{noshx2}). 

\end{itemize}

\section{Proper radial length and a radial coordinate gauge.} 

As mentioned in section 3, the radial coordinate $r$ is not intrinsically covariant. We provide in this Appendix the conditions guaranteeing that the dependence of scalars on $r$ is qualitatively analogous to their dependence on the proper radial length. We also discuss the choice of radial coordinate, given the existence of a radial coordinate gauge freedom in LTB models by virtue of the invariance of the LTB metrics  (\ref{LTB1}) and (\ref{LTB2}) with respect to an arbitrary rescaling $r=r(\bar r)$.

Radial radial rays are spacelike geodesics of the LTB metric \cite{suss10b} whose affine parameter is the proper radial length, which (if there is a symmetry center $r=0$) is given by the function: $\ell[t]:\mathbb{R}^+\to \mathbb{R}$ such that
\begin{equation}\fl \ell[t](r)=\int_{0}^{r}{\sqrt{g_{rr}}\,\dd x}=\int_0^r{\frac{R'}{\sqrt{1+E}}\,\dd x}=\int_0^r{L\,\Gamma\, \frac{R'_i}{\sqrt{1-k_{qi}R_i^2}}\,\dd x}.\label{elldef}\end{equation}
where we used (\ref{ME}) to eliminate $E$ in terms of $k_{qi}$. Since $\ell$ must be a non--negative and monotonously increasing continuous function, the regularity condition (\ref{noshxG}) must hold together with the following extra regularity conditions 
\begin{equation} \frac{R'_i}{\sqrt{1-k_{qi}R_i^2}} > 0,\qquad 1-k_{qi} R_i^2\geq 0,\label{layer}\end{equation}
so that a zero of $R'_i$ (if it exists) must be a common same order zero of $\sqrt{1-k_{qi}R_i^2}$. Since the zeroes of $R'_i$ and $R'$ are common \cite{suss10a,suss10b}, this condition automatically follows if (\ref{noshxG}) holds. The converse is not true, as (\ref{layer}) can be violated even if (\ref{noshxG}) holds. In this latter case, the radial coordinate would be ill defined (as well as the proper radial length), and also a surface layer singularity would arise at the common zero of $R'_i$ and $1-k_{qi}R_i^2$ \cite{ltbstuff, suss02, bonnor85}.

For regular ``open'' models whose $\T[t]$ are topologically equivalent to $\mathbb{R}^3$, or for any LTB region in which $R'>0$ and $1-k_{qi}R_i^2>0$ hold everywhere without violating (\ref{noshxG}) and (\ref{layer}), $R_i$ can be prescribed as any monotonously increasing function complying with $R_i(0)=0$  and $R'_i>0$ for all $r$. Evidently, the simplest choice in these cases is
\begin{equation} R_i = R_0\,r,\label{rgauge}\end{equation}     
where $R_0$ is an arbitrary constant characteristic length scale. The gauge (\ref{rgauge}) is a popular choice in the literature \cite{joshi,num1,num2,LTBfin,LTBchin}, not only due to its simplicity, but because by setting $r=R_i/R_0$ radial dependence becomes dependence on a fiducial value of an invariant quantity that has a clear physical and  geometric meaning ($R$). Also, the choice of $R_0$ provides a physical length scale for the radial coordinate. 

In ``closed'' models whose $\T[t]$ are homeomorphic to $\mathbb{S}^3$ (see section 9) the function $R_i$ cannot be monotonously increasing (and so (\ref{rgauge}) cannot be used). In this case there are two symmetry centers, so that $R_i(0)=R_i(r_c)=0$, hence a ``turning value'' $r=\rtv$ must exist so that $R'_i(\rtv)=R'(ct,\rtv)=0$, but because of (\ref{layer}) $r=\rtv$ must be also a zero of $1-k_{qi}R_i^2$ of the same order. Regular closed models cannot be parabolic or hyperbolic because $k_{qi}\leq 0$ holds for these models, hence (\ref{layer}) would be violated \cite{ltbstuff,suss02,bonnor85}. As a consequence, all regular closed models must be elliptic (and thus collapsing to a second curvature singularity). Since $R'_i$ changes sign at $\rtv$, the regularity condition (\ref{layer}) implies that $\sqrt{1-k_{qi}R_i^2}<0$ must hold in the range $\rtv<r<r_c$ (see \cite{suss10a,suss10b,ltbstuff}).

\section{Initial conditions in terms of an $\Omega$ parameter.}

The fact that $\HH_q$ behaves in (\ref{Hq}) as a Hubble scalar of a FLRW dust universe suggests the definition of a sort of the following ``Omega'' quotient  
\ba\hOm &\equiv& \frac{2m_q}{\HH_q^2}=\frac{2m_q}{2m_q-k_q}=\frac{2m_{qi}}{2m_{qi}-k_{qi}L},\label{Omdef1}
\\\hOm-1 &=& \frac{k_q}{\HH_q^2}=\frac{k_q}{2m_q-k_q} =\frac{k_{qi}L}{2m_{qi}-k_{qi}L}.\label{Omdef2}\ea
This quantity has been used in various articles \cite{num1,LTBfin,suss08} and is only one of the possible generalizations of the Omega parameter for LTB models (see \cite{suss10a} for a proper discussion on this issue). The definitions (\ref{Omdef1})--(\ref{Omdef2}) lead to  
\begin{equation}\fl \hOmi=\frac{2m_{qi}}{\HH_{qi}^2}=\frac{2m_{qi}}{2m_{qi}-k_{qi}},\qquad \hOmi-1=\frac{k_{qi}}{\HH_{qi}^2}=\frac{k_{qi}}{2m_{qi}-k_{qi}},\label{Omidef}\end{equation}
so that the following scaling laws hold:
\begin{equation} \hOm=\frac{\hOmi}{\hOmi-(\hOmi-1)\,L},\qquad \hOm-1=\frac{(\hOmi-1)L}{\hOmi-(\hOmi-1)\,L},\label{Omdef3}\end{equation}
\begin{equation} \HH_q=\HH_{qi}\frac{[\hOmi+(1-\hOmi) L]^{1/2}}{L^{3/2}}.\label{Hqqi}\end{equation}
Notice that prescribing as initial condition a given sign for $\hOmi-1$ completely determines the kinematic class (parabolic, elliptic or hyperbolic) for all times (just as prescribing $k_{qi}$). From (\ref{Omdef2}) and (\ref{Omdef3}) we have:
\ba\hbox{If}\quad \hOmi=1\quad\Rightarrow\qquad \hOm=1,\qquad\hbox{parabolic}\label{Omparab} \\
\hbox{If}\quad 0<\hOmi<1\quad\Rightarrow\quad 0<\hOm<1,\qquad\hbox{hyperbolic}\label{Omhyp}
\\
\hbox{If}\quad\hOmi>1\quad\Rightarrow\qquad \hOm>1,\qquad\hbox{elliptic}\label{Omell}\ea    
Also, it is important to remark that irrespective of the kinematic class we have for every LTB model:\, $\hOm\to 1$ as $L\to 0$ (near a curvature singularity). For all hyperbolic models or regions $\hOmi$ is bounded for all choices of $m_{qi},\,k_{qi}$, hence:\, $\hOm\to 0$ as $L\to\infty$, whereas for elliptic models or regions $\hOm\to \infty$ as $L\to \Lmax$. 

We can always eliminate $m_{qi},\,k_{qi}$ by means of (\ref{Omidef}) in terms of the initial value functions $\hOmi,\,\HH_{qi}$. Since the latter are roughly equivalent to Omega and Hubble factors, or at least they could tend at a given asymptotic limit to these parameters in a FLRW background (see \cite{suss10a} for a discussion on this point), it could be more intuitive to prescribe  $\hOmi,\,\HH_{qi}$ as initial conditions than prescribing $m_{qi},\,k_{qi}$ as in section 4. Since the parabolic case is trivial ($\hOmi=1,\,\HH_{qi}^2=2m_{qi}$), we provide below the hyperbolic and elliptic analytic solutions (\ref{hypZ2}) and (\ref{ellZ2}) rewritten in terms of $\hOmi,\,\HH_{qi}$: 

\begin{itemize}
\item
{\underline{Hyperbolic models or regions: $\hOmi-1\leq 0$}.}
\begin{equation} c(t-t_i)=\frac{W-W_i}{\HH_{qi}}.\label{hypZ3} \end{equation}
\ba 
   \fl  W=\frac{\left[\hOmi+(1-\hOmi)L\right]^{1/2}L^{1/2}}{1-\hOmi}-\frac{\hOmi}{2(1-\hOmi)^{3/2}}\hbox{arccosh}\left(\frac{2L}{\hOmi}+1-2L\right),\label{hypZ4a}\\
   \fl W_i=\frac{1}{1-\hOmi}-\frac{\hOmi}{2(1-\hOmi)^{3/2}}\hbox{arccosh}\left(\frac{2}{\hOmi}-1\right).\label{hypZ4b}\ea

\item
{\underline{Elliptic models or regions: $\hOmi-1\geq 0$}.}
\begin{equation}\fl \HH_{qi}\,c(t - t_i )  = \left\{ \begin{array}{l}
 W-W_i \qquad\qquad\qquad\qquad \hbox{expanding phase}\\ 
  \\ 
 \pi\hOmi(\hOmi-1)^{-3/2} - W-W_i \qquad \hbox{collapsing phase}\\ 
 \end{array} \right.\label{ellZ3}\end{equation}
\ba 
   \fl  W=\frac{\hOmi}{2(\hOmi-1)^{3/2}}\hbox{arccos}\left(\frac{2L}{\hOmi}+1-2L\right)-\frac{\left[\hOmi-(\hOmi-1)L\right]^{1/2}L^{1/2}}{\hOmi-1},\label{ellZ4a}\\
   \fl W_i=\frac{\hOmi}{2(\hOmi-1)^{3/2}}\hbox{arccos}\left(\frac{2}{\hOmi}-1\right)-\frac{1}{\hOmi-1}.\label{ellZ4b}\ea
\end{itemize}

\noindent
Setting $L_i=1,\,\,t=\tbb$ and $L=0$ in these expressiond yields the bang time
\begin{equation}  c\tbb = ct_i-\frac{W_i}{\HH_{qi}},\label{tbbeh}\end{equation}
The time $t=\tmax$ and $t= \tcoll$ in elliptic models, respectively associated with maximal expansion ($L=\Lmax=\hOmi/(\hOmi-1)$) and the collapsing singularity ($L=0$ in the collapsing phase) are
\begin{equation}\fl\tmax=c\tbb+\frac{\pi\hOmi}{2 \HH_{qi} [\hOmi-1]^{3/2}},\qquad  \tcoll=c\tbb+\frac{\pi\hOmi}{\HH_{qi} [\hOmi-1]^{3/2}}.\label{tmcOm}\end{equation}
The relation between $\hOmi,\,\HH_{qi}$ and the conventional free parameters is
\begin{equation} \hOmi=\frac{2M}{2M+E\,R_i},\quad \HH_{qi}=\frac{\left[2M+E\,R_i\right]^{1/2}}{R_i^{3/2}},\label{OmME}\end{equation}
while using (\ref{Dadef}) and (\ref{Omidef}) we can express $\Dim$ and $\Dik$ in terms of $\hOmi,\,\HH_{qi}$ as
\begin{equation} \Dim=\frac{R_i}{3R_i'}\left[\frac{\hOmi'}{\hOmi}+\frac{2\HH_{qi}'}{\HH_{qi}}\right],\qquad \Dik=\frac{R_i}{3R_i'}\left[\frac{\hOmi'}{\hOmi-1}+\frac{2\HH_{qi}'}{\HH_{qi}}\right],\label{OmD}\end{equation}
where in (\ref{OmME}) and (\ref{OmD}) the function $R_i$ can be specified as a choice of radial coordinate gauge.

\section{Local scalar representation.}

Besides $\rho$ given by (\ref{fieldeq2}), other covariant objects associated with LTB models  are the expansion scalar, $\Theta$, the Ricci scalar of the space slices, $\RR$, plus the shear and electric Weyl tensors, $\sigma_{ab},\,E_{ab}$
\ba \fl\Theta &=& \tilde\nabla_au^a=\frac{2\dot R}{R}+\frac{\dot R'}{R'},\qquad
\RR = -\frac{2(E\,R)'}{R^2R'},\label{ThetaRR}\\
\fl \sigma_{ab} &=& \tilde\nabla_{(a}u_{b)}-(\Theta/3)h_{ab}=\Sigma\,\Xi^{ab},\qquad
E^{ab}=  u_cu_d C^{abcd}=\EE\,\Xi^{ab},\label{SigEE}\ea
where \, $\tilde\nabla_a = h_a^b\nabla_b$\, and $C^{abcd}$ is the Weyl tensor, while $\Xi^{ab}=h^{ab}-3\eta^a\eta^b$, with $\eta^a=\sqrt{h^{rr}}\delta^a_r$ being the unit tangent vector along the radial rays (orthogonal to $u^a$ and to the orbits of SO(3)).  The scalars $\EE$ and $\Sigma$ in (\ref{SigEE}) are
\begin{equation}\fl \Sigma = \sigma_{ab}\,\Xi^{ab}=\frac{1}{3}\left[\frac{\dot R}{R}-\frac{\dot R'}{R'}\right],\qquad
\EE = E_{ab}\,\Xi^{ab}=-\frac{\kappa}{6}\,\rho+ \frac{M}{R^3}.\label{SigEE1}\end{equation}
Since LTB models (as all spherically symmetric spacetimes) are LRS (locally rotationally symmetric \cite{LRS}), they can be completely characterized by covariant scalars. Considering (\ref{fieldeq2}), (\ref{ThetaRR}), (\ref{SigEE}) and (\ref{SigEE1}), a choice of scalar representation follows from the local ``fluid flow'' scalars \cite{suss09,suss10a} 
\begin{equation} \{\rho,\,\Theta,\, \RR,\,\Sigma,\, \EE\},\label{locscals}\end{equation}
whose evolution equations completely determine the dynamics of LTB models in the fluid flow or ``1+3'' approach \cite{LRS,ellisbruni89,1plus3}, and thus provide an alternative approach to that based on the analytic solutions of (\ref{fieldeq1}). Our usage of quasi--local scalars provides an alternative complete representation of covariant scalars given by $\{m_q,\,\HH_q,\,k_q,\,\Dm,\,\Dh,\,\Dk\}$ \cite{suss09,suss10a}.

\end{appendix}


\section*{References}             

\begin{thebibliography}{99}


\bibitem{LTB} Lema\^{\i}tre G  1933 {\it Ann. Soc. Sci. Brux.} A  {\bf 53} 51. See reprint in Lema\^{\i}tre G  1997 {\it Gen. Rel. Grav.} {\bf 29} 5;
Tolman R C 1934 {\it Proc. Natl Acad. Sci.} {\bf 20} 169;
Bondi H 1947 {\it Mon. Not. R. Astron. Soc.} {\bf 107} 410.


\bibitem{kras1} Krasi\'nski A, {\textit{Inhomogeneous Cosmological Models}},  Cambridge University Press, 1998.

\bibitem{kras2} Plebanski J and Krasinski A, {\textit{An Introduction to General Relativity and Cosmology}},  Cambridge University Press, 2006.


\bibitem{KH1} Krasi\'nski A and Hellaby C 2002 {\it Phys Rev} D {\bf 65} 023501

\bibitem{KH2} Krasi\'nski A and Hellaby C 2004 {\it Phys Rev} D {\bf 69} 023502

\bibitem{KH3} Krasi\'nski A and Hellaby C 2004 {\it Phys Rev} D {\bf 69} 043502

\bibitem{KH4} Hellaby C and Krasi\'nski A 2006 {\it Phys Rev} D {\bf 73} 023518

\bibitem{BKH} Bolejko K, Krasi\'nski A and Hellaby C 2005 {\it MNRAS} {\bf 362} 213-228


\bibitem{sscoll} Eardley D M 1974 {\it Commun Math Phys} {\bf 37} 287; Eardley D M and Smarr L 1979 {\it Phys Rev} D {\bf 19} 2239; Dyer C C 1979 {\it MNRAS} {\bf 189} 189; Waugh B and Lake K 1988 {\it Phys Rev} D {\bf 38} 1315; Waugh B and Lake K 1989 {\it Phys Rev} D {\bf 40} 2137; Lemos J P S 1991 {\it Phys Lett} A {\bf 158} 279

\bibitem{joshi} Joshi P S and Dwivedi I H 1993 {\it Phys Rev} D {\bf 47} 5357; Joshi P S and Singh T P 1995 {\it Phys Rev} D {\bf 51} 6778; Dwivedi I H and Joshi P S 1997 {\it Class. Quant. Grav.} {\bf 47} 5357


\bibitem{quantum} Vaz C, Witten L and Singh T P 2001 {\it Phys Rev} D {\bf 63} 104020;
Kiefer C, Mueller-Hill, Vaz C 2006 {\it Phys Rev} D {\bf 73} 044025;
Bojowald M, Harada T and Tibrewala R 2008 {\it Phys Rev} D {\bf 78} 064057


\bibitem{LTB1}  Pascual--S\'anchez J F 1999  {\it Mod. Phys. Lett.} A {\bf 14} 1539;  Sugiura N K and Harada T 1999  {\it Phys Rev } D 60 103508; Celeri\`er M N 2000 {\it Astron. Astrophys.} {\bf 353} 63;   Tomita K 2001 {\it MNRAS} {\bf 326} 287;  Iguchi H,  Nakamura T and Nakao K 2002 {\it Prog. Theor. Phys.} {\bf 108} 809; Schwarz D J  2002 Accelerated expansion without dark energy {\it Preprint} {\tt arXiv:astro-ph/0209584v2};    

\bibitem{LTB2}    Apostolopoulos P {\it et al}  2006 {\it JCAP} {\bf P06} 009; Kai T, Kozaki H, Nakao K, Nambu Y and Yoo C M 2007 {\it Prog. Theor. Phys.} {\bf 117} 229-240 ({\it Preprint} {\tt  arXiv:gr-qc/0605120});   Mattsson T and Ronkainen M 2008  {\it JCAP} {\bf 0802} 004 ({\it Preprint} {\tt arXiv:astro-ph/0708.3673v2});   Bolejko K and Andersson L 2008 {\it JCAP} {\bf 10} 003 ({\it Preprint} {\tt arXiv:0807.3577})

\bibitem{LTBkolb} Kolb E W, Matarrese S, Notari A and  Riotto A 2005 {\it Phys Rev} D {\bf 71} 023524  ({\it Preprint} {\tt arXiv:hep-ph/0409038v2}); Marra V,  Kolb E W and Matarrese S 2008 {\it Phys Rev} D {\bf 77} 023003; Marra V, Kolb E W, Matarrese S and Riotto A 2007 {\it Phys Rev} D {\bf 76} 123004. 

\bibitem{num1} Garc\'\i a--Bellido J and Troels H 2008 {\it J. Cosmol. Astropart. Phys.} {\bf JCAP} 0804:003 {\tt Preprint gr-qc/0802.1523v3 [astro-ph]}

\bibitem{num2} Moffat J W 2006 {\it J. Cosmol. Astropart. Phys.} {\bf JCAP} (2006)001; Alnes H, Amazguioui M and Gron O 2006 {\it Phys Rev} D {\bf 73} 083519; Alnes H and Amazguioui M 2006 {\it Phys Rev} D {\bf 74} 103520;  Alnes H and Amazguioui M 2006 {\it Phys Rev} D {\bf 75} 023506


\bibitem{LTBfin}   Rasanen S 2006 {\it Class. Quant. Grav.} {\bf 23} 1823-1835;   Enqvist K and  Mattsson T 2007 {\it JCAP} {\bf 0702} 019 ({\it Preprint} {\tt 	arXiv:astro-ph/0609120v4}); Enqvist K 2008 {\it Gen. Rel. Grav.}  {\bf 40}  451-466 (Preprint {\tt arXiv:0709.2044})

\bibitem{LTBchin} Chuang C H, Gu J A and Hwang W Y P 2005 {\it Class.Quant.Grav.},{\bf 25}, 175001 {\tt Preprint  astro-ph/0512651}

\bibitem{LTBave1} Paranjape A and Singh T P 2006 {\it Class.Quant.Grav.},{\bf 23}, 6955Ð6969

\bibitem{LTBave2} Sussman R A 2008 On spatial volume averaging in Lema"tre--Tolman--Bondi dust models. Part I: back reaction, spacial curvature and binding energy  {\it Preprint} {\tt arXiv:0807.1145 }

\bibitem{LTBave3} Sussman R A 2009 Quasi-local variables and scalar averaging in LTB dust models {\it Preprint} {\tt arXiv:0912.4074 } 


\bibitem{celerier} Celeri\`er M N 2007 {\it New Advances in Physics} {\bf 1} 29 ({\it Preprint} {\tt arXiv:astro-ph/0702416})

\bibitem{ave_review} Buchert T  2000 {\it Gen. Rel. Grav}  {\bf 9}  306-321 ({\it Preprint} {\tt arXiv:gr-qc/0001056v1}); Buchert T 2008 {\it Gen. Rel. Grav.} {\bf 40}, 467

\bibitem{wainwright} Wainwright J and Andrews S 2009 {\it Class.Quant.Grav.},{\bf 26}, 085017


\bibitem{CBK} Celeri\`er M N Bolejko K and Krasinski A 2009 A (giant) void is not mandatory to explain away dark energy with a Lema\^\i tre--Tolman model {\it Preprint} {\tt arXiv:0906.0905 }


\bibitem{mushel} Mustapha N and Hellaby C 2001 {\it Gen Rel Gravit} {\bf 33} 455


\bibitem{occhio81} Occhionero F Veccia--Scavalli L and Vittorio N 1981 {\it Astron Astrophys} {\bf 97} 169

\bibitem{occhio83} Occhionero F Santangelo P and Vittorio N 1983 {\it Astron Astrophys} {\bf 117} 365

\bibitem{cham} Chamorro A 1991 {\it Astrophys J} {\bf 383} 51

\bibitem{sato} Maeda K Sasaki M and Sato H 1983 {\it Progr Theo Phys} {\bf 69} 89 

\bibitem{boncham1} Bonnor W B and Chamorro A 1990 {\it Astrophys J} {\bf 361} 21

\bibitem{boncham2} Bonnor W B and Chamorro A 1991 {\it Astrophys J} {\bf 378} 461

\bibitem{meszaros} Meszaros A 1993 {\it Astrophys Space Sci} {\bf 207} 5


\bibitem{sussQL1} Sussman R A 2008 Quasi-local variables, non-linear perturbations and back-reaction in spherically symmetric spacetimes {\it Preprint} {\tt arXiv:0809.3314}

\bibitem{sussQL} Sussman R A Quasi-local variables and inhomogeneous cosmological sources with spherical symmetry 2008 {\it AIP Conf.Proc.} {\bf 1083} 228-235 {\tt Preprint arXiv:0810.1120}.

\bibitem{suss08}Sussman R A 2008 {\it Class Quantum Grav.}  {\bf 25} 015012 {\tt Preprint arXiv:grÐqc/0709.1005}

\bibitem{suss09} Sussman R A 2009 {\it Phys Rev} D {\bf 79} 025009. 

\bibitem{suss10a} Sussman R A 2010 A new approach for doing theoretical and numeric work with Lema\^{\i}tre--Tolman--Bondi dust models {\it Preprint} {\tt arXiv:1001.0904v1 }

\bibitem{suss10b} Sussman R A 2010  Radial asymptotics of Lema\^\i tre--Tolman--Bondi dust models {\it Preprint} {\tt arXiv:1002.0173 }



\bibitem{HLconds} Hellaby C and Lake K 1985 {\it Astrophys J.} {\bf 290} 381

\bibitem{ltbstuff} Matravers D R and Humphreys N P  2001 {\it Gen. Rel. Grav.} {\bf 33} 
531Ð52; Humphreys N P, Maartens R and Matravers D R 1998 Regular spherical dust spacetimes {\tt Preprint gr-qc/9804023v1}.


\bibitem{LRS} van Elst H and  Ellis G F R 1996 {\it Class Quantum Grav} {\bf 13} 1099-1128 ({\it Preprint} {\tt  arXiv:gr-qc/9510044})

\bibitem{ellisbruni89} Ellis G F R and Bruni M 1989 {\it Phys Rev} D {\bf 40}  1804

\bibitem{1plus3} Ellis G F R and van Elst H 1998 Cosmological Models (Carg\`ese Lectures 1998) {\it Preprint} {\tt arXiv gr-qc/9812046 v4}


\bibitem{hayward} Hayward S A 1996 {\it Phys Rev} D {\bf 53} 1938 ({\it Preprint} {\tt ArXiv gr-qc/9408002}); Hayward S A 1998 {\it Class Quantum Grav} {\bf 15} 3147Ð3162  ({\it Preprint} {\tt ArXiv gr-qc/9710089v2})


\bibitem{suss02} Sussman R A and  Garc\'\i a--Trujillo L 2002 {\it Class.Quant.Grav.}  {\bf 19} 2897-2925.

\bibitem{bonnor85} Bonnor W B {\it Class Quantum Grav} {\bf 2} 781--790


\bibitem{padma} Padmanabhan T, {\textit{Theoretical Astrophysics Volume III: Galaxies and Comsology}},  Cambridge University Press, 2002. See chapter 5.10 and figure 5.3.





\bibitem{wiltshire} Wiltshire D 2007 {\it New J. Physics} {\bf 9} 377  ({\it Preprint} {\tt ArXiv gr-qc/0702082v4})

\bibitem{fiEllis} Ellis G F R, in Bertotti B de Felice F and Pacolini A (eds), {\it{General Relativity and Gravitation}} (Reidel, Dordrecht, 1984) pp 215--288; 










\end{thebibliography}
\end{document}